\documentclass[sn-mathphys,Numbered]{sn-jnl}% Math and Physical Sciences Reference Style
\usepackage{graphicx}%
\usepackage{multirow}%
\usepackage{amsmath,amssymb,amsfonts}%
\usepackage{amsthm}%
\usepackage{mathrsfs}%
\usepackage[title]{appendix}%
\usepackage{xcolor}%
\usepackage{textcomp}%
\usepackage{manyfoot}%
\usepackage{booktabs}%
\usepackage{algorithm}%
\usepackage{algorithmicx}%
\usepackage{algpseudocode}%
\usepackage{listings}%
\theoremstyle{thmstyleone}%
\newtheorem{theorem}{Theorem}%  meant for continuous numbers
\newtheorem{proposition}[theorem]{Proposition}% 

\theoremstyle{thmstyletwo}%

\theoremstyle{thmstylethree}%

\usepackage{amsmath}
\usepackage{amsfonts}
\usepackage{amssymb}
\usepackage{subcaption}
\newcommand{\E}{\mathcal{E}}
\begin{document}

\title[Forward hysteresis and Hopf bifurcation in a NPZD model]{Forward hysteresis and Hopf bifurcation in an NPZD model with application to harmful algal blooms}


\author*[1,2]{\fnm{J. C.} \sur{Macdonald}}\email{joshuamac@tauex.tau.ac.il}

\author*[1]{\fnm{H.} \sur{Gulbudak}}\email{hayriye.gulbudak@louisiana.edu}


\affil[1]{\orgdiv{Department of Mathematics}, \orgname{University of Louisiana at Lafayette}, \orgaddress{\street{1401 Johnston Street}, \city{Lafayette}, \postcode{70504}, \state{Louisiana}, \country{USA}}}
\affil[2]{\orgdiv{Current address: School of Zoology, Faculty of Life Sciences}, \orgname{Tel Aviv University}, \city{Tel Aviv-Yafo}, \country{Israel}}
% \affil[2]{\orgdiv{Department}, \orgname{Organization}, \orgaddress{\street{Street}, \city{City}, \postcode{10587}, \state{State}, \country{Country}}}

% \affil[3]{\orgdiv{Department}, \orgname{Organization}, \orgaddress{\street{Street}, \city{City}, \postcode{610101}, \state{State}, \country{Country}}}


\abstract{Nutrient-Phytoplankton-Zooplankton-Detritus (NPZD) models, describing the interactions between phytoplankton, zooplankton systems, and their ecosystem, are used to predict their ecological and evolutionary population dynamics.  These organisms form the base two trophic levels of aquatic ecosystems. Hence understanding their population dynamics and how disturbances can affect these systems is crucial. Here, starting from a base NPZ modeling framework, we incorporate the harmful effects of phytoplankton overpopulation on zooplankton - representing a crucial next step in harmful algal bloom (HAB) modeling - and split the nutrient compartment to formulate an NPZD model.  We then mathematically analyze the NPZ system upon which this new model is based, including local and global stability of equilibria, Hopf bifurcation condition, and forward hysteresis, where the bi-stability occurs with multiple attractors.  Finally, we extend the threshold analysis to the NPZD model, which displays both forward hysteresis with bi-stability and Hopf bifurcation under different parameter regimes, and examine ecological implications after incorporating seasonality and ecological disturbances. Ultimately, we quantify ecosystem health in terms of the relative values of the robust persistence thresholds for phytoplankton and zooplankton and find (i) ecosystems sufficiently favoring phytoplankton, as quantified by the relative values of the plankton persistence numbers, are vulnerable to both HABs and (local) zooplankton extinction (ii) even healthy ecosystems are extremely sensitive to nutrient depletion over relatively short time scales.
}


\maketitle

\section{Introduction}\label{sec1}

\label{intro}

Plankton are a ubiquitous group of very small drifting organisms which live in both salty and fresh water and form the base trophic levels of aquatic food webs \cite{karleskint2012introduction}.  Divided by trophic position, phytoplankton are primary producers, generating growth through photosynthesis, while zooplankton are primary consumers, and feed primarily on phytoplankton \cite{karleskint2012introduction}. Mixotrophic organisms, those which can opportunistically switch between photosynthesis and heterotrophy, also exist in this group \cite{tittel2003mixotrophs, flynn2013misuse}. 
  Disturbance is an important mechanism that affects the functioning of these ecosystems, and variation in type, frequency, intensity, and duration of disturbance has important implications for ecosystem and community structure and thus underlying population dynamics \cite{hobbs1992disturbance}.  There is general agreement that a high frequency of ecological disturbance has a net negative effect on ecosystem species diversity \cite{fox2013intermediate}.  Given plankton's foundational position in aquatic food webs, understanding their interactions and population dynamics is a central focus in aquatic ecology (cf \cite{richardson, barton, chust}) - not least because they can be main drivers of knock-on effects of ecological disturbances.  Here we focus in particular on plankton population dynamics in the wake of nutrient influx (eutrophication, \cite{hallegraeff1993review,anderson}) and nutrient depletion events (re-oligotrophication, see  \cite{opposite}), which are of immediate concern given their potential cascading cross-scale knock-on effects which can range from local die-offs \cite{trubovitz2020marine}, to diverse disease spillover events and pathogenesis of vector-borne disease (particularly in freshwater ecosystems, \cite{hilborn2014algal, kouakou2019economic,lassudrie2020effects}), and that under the current climate regime, conditions that promote their occurrence have been both predicted to and shown to increase (\cite{anderson,hallegraeff1993review,hu,opposite}) in contrast to other potential disturbances. 
  
Plankton blooms are naturally occurring in temperate zones during the spring \cite{karleskint2012introduction}.  However, phytoplankton blooms may cause deleterious ecological effects through toxicity (e.g., harmful algal blooms (HABs) or through the formation of low oxygen/hypoxic zones, cf \cite{anderson2001nutrient,diaz2001overview,sylvan2006phosphorus}), both of which negatively affect ecological consumers (e.g., zooplankton or fishes;  \cite{anderson,hallegraeff1993review,ralston2020modeling}).  There is debate about what exactly is classified as a bloom versus what is not (but see \cite{bloomdef} for one definition).  Here, we take a conservative approach and say a bloom has occurred if the peak phytoplankton concentration is at least $300\%$ the mean value of the time simulation over the course of one year, and investigate their effects as a disturbance of plankton population dynamics.

Nutrient-Phytoplankton-Zooplankton-Detritus (NPZD)  models (cf. \cite{mackas, busenberg,henderson,edwards,edwards2001adding,franks,heinle,kloosterman}) have long been used by plankton ecologists to investigate plankton interactions and long-term population dynamics in various settings from lakes to the far-from-shore ocean.  These models are essentially nested Lotka-Volterra models with phytoplankton `predating' nutrients and zooplankton predating phytoplankton and are used to study the mechanisms that sustain plankton coexistence and diversity \cite{franks,edwards2001adding}. 
 Hopf bifurcation, which biologically corresponds to predator-prey feedback loops, is a common feature of these models (cf. \cite{busenberg,edwards}), and forward hysteresis has been demonstrated depending on the choice of zooplankton mortality term \cite{edwards}.  Nutrient cycling has also been modeled by incorporating delay differential equations instead of a separate detritus compartment \cite{kloosterman}. However, many of these models - particularly those with many nonlinear terms - have not been mathematically analyzed, which is crucial for understanding complex interactions between NPZ(D) systems, and have not been used in the context of ecological disturbances. 
 
In this study, starting from a base NPZ modeling framework from the literature \cite{henderson,edwards}, we incorporate the harmful effects of phytoplankton overpopulation on zooplankton by adding a functional form to the zooplankton compartment and split the nutrient compartment to formulate an NPZD model (Section \ref{sec:incorp}).  To our best knowledge, this is the first multi-trophic level model to incorporate process-based effects of harmful algal blooms (HABs), which has been argued for as a crucial next step in HAB modeling \cite{ralston2020modeling}.   We then mathematically analyze the NPZ system upon which this new model is based with both quadratic and linear zooplankton mortality terms (Section \ref{sec:3}) - deriving global stability conditions for the zooplankton extinction equilibrium in terms of zooplankton invasion number for a special case of the linear mortality model.  We also derive local stability, Hopf bifurcation, and derive existence conditions for the coexistence equilibria of the NPZ model with both quadratic and linear zooplankton mortality.  We also provide one and two-parameter bifurcation diagrams for both models, showing forward hysteresis with bi-stability or Hopf bifurcation in the quadratic loss case depending on parameter values, along with Hopf bifurcation or transcritical bifurcation in the linear loss case.  Finally, we extend the threshold analysis to the new NPZD model, which displays both forward hysteresis with bi-stability and Hopf bifurcation, and examine ecological implications after incorporating seasonality and ecological disturbances in the form of eutrophication and re-oligotrophication events (Section \ref{sec:eco_imp}).  Ultimately we quantify ecosystem health in terms of the relative values of the robust persistence thresholds for phytoplankton and zooplankton and find (i) ecosystems sufficiently favoring phytoplankton are vulnerable to both HABs and (local) zooplankton extinction (ii) even balanced ecosystems are extremely sensitive to nutrient depletion over relatively short time scales.

\section{Modeling NPZ and NPZD systems}
\label{sec:incorp}
The general form of the coupled nonlinear NPZ ODE models describes the interactions between nutrients (N), phytoplankton (P), and zooplankton (Z).
The terms $f(P), g(N)$ represent phytoplankton's response to light (ie. growth), and phytoplankton nutrient uptake, respectively. The terms $h(P), i(P)$ denote phytoplankton mortality due to zooplankton grazing, and natural mortality, respectively.
 Zooplankton growth is driven by interaction with phytoplankton, scaled by zooplankton assimilation, which can be thought of as `messy eating'.  Because zooplankton are the highest trophic level modeled, all zooplankton loss is generally accounted for with a single functional form, $j(Z)$.  In addition to nutrient loss due to phytoplankton nutrient uptake, nutrient loss/exchange can also occur due to other factors such as re-oligotrophication \cite{opposite} or cross-thermocline exchange of nutrients \cite{mackas}, which is represented by the functional term $m(N)$. Furthermore, nutrient growth results from plankton death and inefficient zooplankton grazing, denoted with the term $(1 - \gamma)h(P)Z$, where $\gamma$ represents the zooplankton assimilation rate. The general form of the NPZ model is given as follows \cite{franks} (see schematic Figure \ref{fig:Diagram}):

\begin{figure}[h!]
    \centering
    \includegraphics[width= \textwidth]{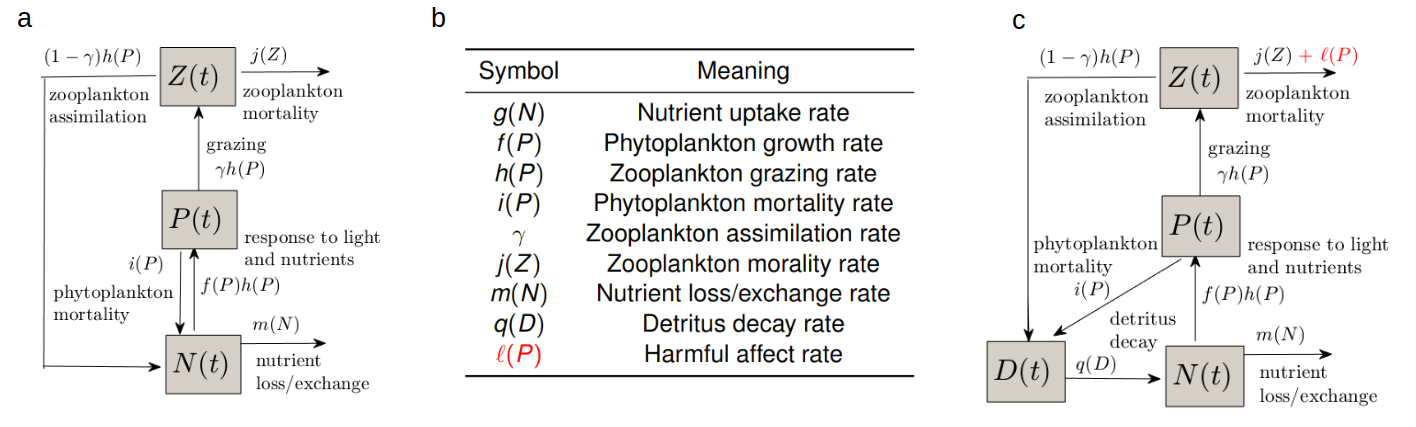}
    \caption{\textbf{The general NPZ and NPZD modeling framework} \textbf{(a)} The general NPZ model \textbf{(b)} Table with a description of model transition rates.  \textbf{(c)} The general NPZD model with new functional form, $\ell(P)$, highlighted in red. }
    \label{fig:Diagram}
\end{figure}

%Functional forms are generally continuous and non-negative in $\mathbb{R_+}$ \cite{franks}.  

% \begin{table}[h!]
% \caption{\label{tab:fxnlforms}Summary of functional forms in the system (1)}
% \centering
% \begin{tabular}{cc}
% \hline\noalign{\smallskip}
% Symbol & Meaning \\
% \noalign{\smallskip}\hline\noalign{\smallskip}
% $f(I)$ & phytoplankton response to light \\

% $g(N)$ & phytoplankton nutrient uptake \\

% $h(P)$ & zooplankton grazing \\

% $i(P)$ & phytoplankton loss due to various factors \\

% $\gamma$ & zooplankton assimilation \\
% $j(Z)$ & zooplankton loss due to various factors. \\
% $m(N)$ & nutrient loss due to various factors. \\
% \noalign{\smallskip}\hline
% \end{tabular}
% \end{table}
\begin{equation}\label{model1}
\begin{cases}
\dot{P}=f(P)g(N)P - h(P)Z - i(P)P,  \\

\dot{Z} = \gamma h(P)Z - j(Z)Z,  \\

\dot{N} = -f(P)g(N)P + (1 - \gamma)h(P)Z + i(P)P  - m(N).
\end{cases} 
\end{equation}
Following \cite{,henderson,edwards}, we consider the specific functional forms:
\begin{equation}
\begin{split}
    f(P)&=\beta\left(1 - \frac{P}{K}\right),\ g(N)=\frac{N}{k + N},\ h(P)= \frac{\eta P^2}{\mu^2 + P^2}, \\ j(Z)&=\delta Z^\sigma,i(P)=0, m(N) = S(N-\Theta).
\end{split}
\end{equation}  
Therefore one can obtain the following coupled ODE model:
\begin{equation}
\left\lbrace
\begin{split}
\dot{P} &=  \frac{N}{k + N}\beta P\left(1 - \frac{P}{K}\right) - \frac{\eta P^2}{\mu^2 + P^2}Z, \\
\dot{Z} &= \gamma\left[\frac{\eta P^2}{\mu^2 + P^2} - \delta Z^{\sigma}\right]Z,  \\
\dot{N} &= -\beta \frac{N}{k + N}\left(1 - \frac{P}{K}\right)P + (1 - \gamma)\frac{\eta P^2}{\mu^2 + P^2}Z  - S(N - \Theta)
\end{split}
\right.
\label{nonRescale1}
\end{equation}

In the NPZ model \eqref{nonRescale1}, the modified logistic equation for phytoplankton growth has the saturating response of Holling's type II for nutrient uptake (Michaelis-Menten kinetics). We also have a saturating response of Holling's type III function for the zooplankton grazing of phytoplankton \cite{holling}.  These are chosen because plankton are known to not have well-mixed spatial distributions \cite{mackas}. The parameters $k, K, \eta$ and $\mu$ denote the Michaelis-Menton half saturation constant, the phytoplankton carrying capacity, the maximum zooplankton grazing rate, and the half-saturation grazing constant, respectively. Zooplankton growth is in direct response to interaction with phytoplankton, scaled by the zooplankton assimilation rate, $\gamma$, and loss is assumed to be either linear ($\sigma = 0$) or quadratic ($\sigma = 1$) with death rate $\delta$.  
%Nutrient growth corresponds to plankton mortality, and nutrient loss to phytoplankton growth as well as various environmental factors, where 
Moreover, the parameter $S$ represents the nutrient loss/exchange rate. Finally, the parameter $\Theta$ represents the intrinsic nutrient level. %The highest trophic level explicitly modeled is zooplankton.  

\subsection{Incorporating harmful affect of phytoplankton overpopulation}
To account for the harmful effect of phytoplankton overpopulation on zooplankton during harmful algal blooms, as well as nutrient cycling, we incorporate a new variable, $D(t)$, representing the amount of Detritus, to the model \eqref{model1}, providing a general NPZD framework (see Fig.\ref{fig:Diagram}c)
We then define the net reproductive rate of zooplankton as a function of phytoplankton as follows:  
\begin{equation}\label{rP}
    r(P) =  \underbrace{h(P)}_{\textrm{zooplankton grazing}} - \underbrace{\ell(P)}_{\textrm{harmful affect of phytoplankton}},
\end{equation}
where $\ell(P) = \frac{\alpha P}{\Xi + P}$  with $\alpha$ representing the maximum harmful effect of phytoplankton on zooplankton, and $\Xi$ denoting the half-saturation constant of the effect. The term $\ell(P)Z$ represents the deleterious effect of phytoplankton overpopulation on zooplankton. Additionally, to better capture nutrient cycling we split the compartment, $\dot{N},$ into two compartments, with $\dot{D}$ representing the rate of change in detritus concentration which decays into nutrients at rate $q(D)=\Psi$, and phytoplankton having natural mortality rate $i(P) = \epsilon$.  

\begin{figure}[h!]
    \centering
  \begin{subfigure}[t]{.45\textwidth}
     \includegraphics[width=\textwidth]{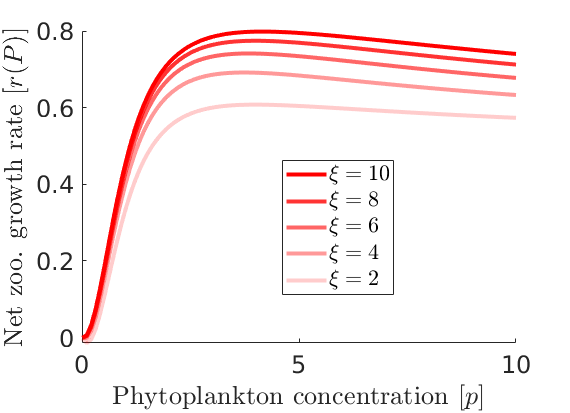}
           \caption{}
     \end{subfigure}
     \begin{subfigure}[t]{.45\textwidth}
     \includegraphics[width=\textwidth]{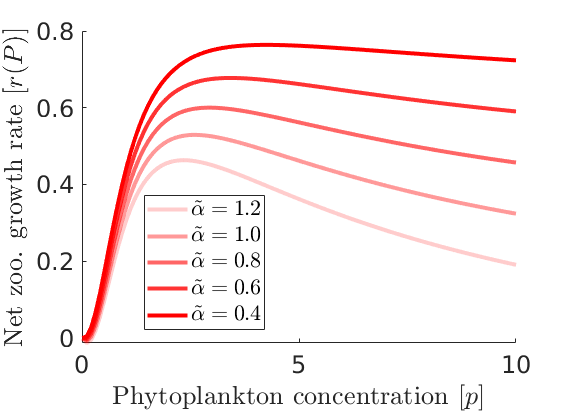}
         \caption{}
     \end{subfigure}
    \caption{{\bf The net zooplankton reproduction rate as a function of phytoplankton population}. The precise shape of $r(P)$ is dictated by two rescaled model parameters: $\xi$ and $\tilde{\alpha}$ \eqref{rescale}, which represent the half-saturation constant and maximum harmful effect of phytoplankton on zooplankton respectively.  In panel \textbf{(a)}, the net zooplankton growth rate increases with $\xi$ ($\tilde{\alpha}$ is fixed at 0.5), and  in panel \textbf{(b)}, it decreases as $\tilde{\alpha}$ increases ($\xi$ is fixed at 5).}
    \label{fig:netgrowth}
\end{figure}

This functional form of $r(P),$ given in  \eqref{rP}, represents the beneficial effect of phytoplankton grazing for zooplankton population, and the deleterious effect of phytoplankton overpopulation on zooplankton.
This formulation is motivated by our prior work, modeling antibody-dependent enhancement (ADE) in Dengue,  where an increase in preexisting crossreactive antibodies can increase infection severity \cite{gulbudak2020infection}. Here we assume that the net zooplankton reproduction can decrease as the phytoplankton population size $P$ achieves a peak above some threshold, representing harmful bloom occurrence \cite{anderson,hallegraeff1993review} (see figure \ref{fig:netgrowth}).
%We assume simple exponential terms to describe the natural mortality of phytoplankton and the decay of detritus into nutrients.  
Taken together this results in the model
\begin{equation}
\left\lbrace
\begin{split}
\dot{P} &=  \frac{N}{k + N}\beta P\left(1 - \frac{P}{K}\right) - \frac{\eta P^2}{\mu^2 + P^2}Z - \epsilon P,\\
\dot{Z} &= \gamma\left(\left[\frac{\eta P^2}{\mu^2 + P^2} - \frac{\alpha P}{\Xi + P} \right]Z - \delta Z^2\right),  \\
\dot{N} &= -\frac{N}{k + N}\beta P\left(1 - \frac{P}{K}\right)   - S(N - \theta) + \Psi D \\
\dot{D} &= (1 - \gamma)\frac{\eta P^2}{\mu^2 + P^2}Z  + \epsilon P - \Psi D. 
\end{split}
\right.
\label{nonrescale}
\end{equation}

This model can be reparamatrized via
\begin{equation}
\left\lbrace
\begin{split}
\tau &= \beta t, p = \mu^{-1}P, z = \eta (\beta\mu)^{-1}Z,n= \mu^{-1}N, \\
a &= \delta\beta\mu/\eta^{2},s = S\beta^{-1},\theta = \Theta\mu^{-1}, \\ 
\tilde{k} &= k\mu^{-1},c = K\mu^{-1}, \tilde{\gamma} = \eta\gamma\beta^{-1},
\end{split}\right.
\end{equation} 
similar to the NPZ models in \cite{edwards,henderson}.
%because we wish to keep comparability for simulations of model \eqref{rescale} to the classic model with quadratic mortality, and because these choices are biologically justified. 
To further rescale new parameters in the model \eqref{nonrescale},  we consider 
\begin{equation}
\left\lbrace\begin{split}
    \tilde{\alpha} &= \alpha\eta^{-1}, \xi = \Xi\mu^{-1}, d = D\mu^{-1} \\ \psi &= \Psi\beta^{-1}, \tilde{\epsilon} = \epsilon\beta^{-1},
\end{split}\right.
\end{equation}
%resulting in the system \eqref{rescale}. 

% \begin{equation}
% \left\lbrace
% \begin{split}
% \frac{dp}{d\tau} &= \frac{n}{\tilde{k} + n}p\left(1 - \frac{p}{c}\right) - \frac{p^2}{1 + p^2}z \\
% \frac{dz}{d\tau} &= \tilde{\gamma}\left[\frac{p^2}{1 + p^2} - az^\sigma\right]z \\
% \frac{dn}{d\tau} &= -\frac{n}{\tilde{k} + n}p\left(1 - \frac{p}{c}\right) + (1 - \gamma)\frac{p^2}{1 + p^2}z + s(\theta - n)
% \end{split}
% \right.
% \label{rescale}
% \end{equation}

Then the full NPZD model \eqref{nonrescale} can be re-scaled as:

\begin{equation}
\left\lbrace
\begin{split}
\frac{dp}{d\tau} &= \frac{n}{\tilde{k} + n}p\left(1 - \frac{p}{c}\right) - \frac{p^2}{1 + p^2}z - \tilde{\epsilon}p\\
\frac{dz}{d\tau} &= \tilde{\gamma}\left[\frac{p^2}{1 + p^2} -\frac{\tilde{\alpha}p}{\xi+p} - az\right]z \\
\frac{dn}{d\tau} &= -\frac{n}{\tilde{k} + n}p\left(1 - \frac{p}{c}\right)  - s(n - \theta) + \psi d \\
\frac{dd}{d\tau} &= (1 - \gamma)\frac{p^2}{1 + p^2}z + \tilde{\epsilon}p - \psi d
\end{split}
\right.
\label{rescale}
\end{equation}
%It is this model and modifications thereof we use for mathematical analysis, simulations, and bifurcation diagrams (see the SI).

In the system \eqref{rescale}, the parameters $a,s$ are respectively the rescaled zooplankton mortality and nutrient loss/exchange rates.  Phytoplankton carrying capacity is represented by $c$, and $\tilde{\gamma}$ represents zooplankton assimilation scaled by the ratio of phytoplankton response to light, $\beta$, and maximum zooplankton grazing rate $\eta$.

\section{Threshold analysis of the NPZ model}
\label{sec:3}
To understand the model dynamics and ecological impact of HABs, we first analyze the NPZ system \eqref{nonRescale1} then we extend the analysis to the NPZD model \eqref{nonrescale} along with crucial ecological implications.

Note that the rescaled NPZ subsystem \eqref{nonRescale1} is as follows \cite{edwards}:
%All subsequent mathematical analysis (Sec. \ref{sec:3}), bifurcation diagrams (Figs. \ref{fig:linear},\ref{fig:quadratic}, Tab. \ref{tab:sims}), simulations (Figs. \ref{fig:linear}-\ref{fig:bloomLinear}, Tab. \ref{tab:sims}), and discussion of ecological implications (Secs. \ref{sec:4},\ref{sec:5}, Tab. \ref{tab:sims}) are for this system.

\begin{equation}
\left\lbrace
\begin{split}
\frac{dp}{d\tau} &= \frac{n}{\tilde{k} + n}p\left(1 - \frac{p}{c}\right) - \frac{p^2}{1 + p^2}z \\
\frac{dz}{d\tau} &= \tilde{\gamma}\left[\frac{p^2}{1 + p^2} - az^\sigma\right]z \\
\frac{dn}{d\tau} &= -\frac{n}{\tilde{k} + n}p\left(1 - \frac{p}{c}\right) + (1 - \gamma)\frac{p^2}{1 + p^2}z - s(n - \theta)
\end{split}
\right.
\label{rescale2}
\end{equation}
For the analysis of the model \eqref{rescale2}, we consider two cases: (i) quadratic zooplankton loss term $(\sigma = 1)$ and (ii) linear zooplankton loss term $(\sigma = 0).$  All model parameters are assumed to be strictly positive. In both the case of a quadratic zooplankton loss term $(\sigma = 1)$ and a linear zooplankton loss term $(\sigma = 0)$, the model has two boundary equilibria in the positive phytoplankton-nutrient plane: $\E_{np} = (c,0,\theta)$ and $\E_{n} = (0,0,\theta)$.  

For both of these equilibria, the steady state nutrient level is the intrinsic nutrient level of the system, $\theta$.  Equilibrium $\E_{np}$ corresponds to the steady state with phytoplankton at their carrying capacity, $c$, and equilibrium $\E_{n}$ represents community collapse.  As we show below, changes in zooplankton loss term induce distinct qualitative dynamics.  Regardless of the choice of zooplankton mortality functional form, the community collapse equilibrium is always unstable for this model.  

\begin{proposition}
The community collapse equilibrium, $\E_{n}$, of model \eqref{rescale2} is always unstable. 
\label{prop:append}
\end{proposition}
\begin{proof}
See appendix A.
\end{proof}

See Table \ref{tab:threshold} for a summary of results.

% \begin{theorem}  The system (3) is well-posed and dissipative.\end{theorem} 
% \begin{proof}
% See appendix.
% \end{proof}

\subsection{Linear zooplankton loss term ($\sigma = 0$) with the full model}
\label{sec:3a}

% \begin{proposition}
% When $\sigma = 0$ the community collapse equilibrium, $\E_n = (c,0,\theta)^T$, is always unstable.
% \end{proposition}
% \begin{proof}
% See appendix. 
% \end{proof}
% \begin{proof}
% \textcolor{red}{move to appendix} Consider the Jacobian matrix evaluated at this point, which is \begin{equation}
%     J(\E_n = \begin{pmatrix}
% 		\frac{n_i}{\tilde{k} + n_i} & 0 & 0\\
% 		\\
% 		0 & -\tilde{\gamma}a & 0 \\ \\
% 		-\frac{n_i}{\tilde{k} + n_i} & 0 & -s
% 	\end{pmatrix}.
% \end{equation}
% From the form of (4) it is readily apparent that $\E_n$ is a hyperbolic saddle node.
% \end{proof}
Define the zooplankton invasion number, \begin{equation}\mathcal{R}_0:= \frac{1}{a}\cdot\frac{c^2}{1+c^2}. \label{r0}\end{equation}

% \begin{proposition}
% When $\sigma = 0$ the zooplankton extinction equilibrium, $\E_{np} = (c,0,\theta)^T$, is locally asymptotically stable if invasion number,$
%     \mathcal{R}_0 < 1$.  It is unstable otherwise.
% \end{proposition}
% \begin{proof}
% For the boundary equilibrium $\E_{np}$ we have the Jacobian:
%   \begin{equation}
%     J\vert_{\E_{np}} = \begin{pmatrix} A & 0 \\ * & - s \end{pmatrix},
% \end{equation}
% where $$A = 
% \begin{pmatrix}
% -\frac{n_i}{k+n_i} & -a\mathcal{R}_0 \\
% & \\
% 0 & \tilde{\gamma}a(\mathcal{R}_0 - 1)
% \end{pmatrix},$$
% is block triangular, and $\lambda_z^{\E_{np}} = \tilde{\gamma}a(\mathcal{R}_0 - 1)$.  From this it is clear that this eigenvalue is strictly positive if $\mathcal{R}_0 > 1$.
% \end{proof}
% \begin{proof}
% See appendix..
% \end{proof}

% \noindent In Prop. 3 the following result of Barbashin and Krasovskii \cite{barbashin} is used: \\

% \noindent\textbf{Theorem} (global asymptotic stability) for equilibrium $\mathbf{\bar{x}}$, and Lyapunov function $V(\mathbf{x})$ satisfying 
% \begin{enumerate}
%     \item $\dot{V} < 0$ for $\mathbf{x} \neq \mathbf{\bar{x}}$ \hspace{.5in}($V$ is a strict Lyapunov Function)
%     \item $\lim_{\vert\vert\mathbf{x}\vert\vert\rightarrow \infty}V = \infty$ \hspace{.5in}($V$ is radially unbounded)
% \end{enumerate}
% in some set $A \subseteq \mathbb{R}^n$, then
% it follows that $\mathbf{\bar{x}}$ is globally asymptotically stable in $A$.

\begin{proposition}
If $\mathcal{R}_0 < 1$ and $\sigma = 0$, then the zooplankton extinction equilibrium of system \eqref{rescale2}, $\E_{np}$, is locally asymptotically stable.  If $\mathcal{R}_0 > 1$, it is unstable.
\label{prop:1}
\end{proposition}
\begin{proof}
See appendix A.
\end{proof}
From expression \eqref{r0} it is apparent that the invasion number $\mathcal{R}_0$ may be interpreted as the average lifespan of zooplankton, $a^{-1}$, scaled down by a function of the phytoplankton carrying capacity, $c^2(1+c^2)^{-1}$.  Note that for fixed $a$, $\lim_{c\rightarrow\infty}R_0 = a^{-1}$ and $\lim_{c\rightarrow 0}R_0 = 0$.  Similarly, for fixed $c$, $\lim_{a\rightarrow 0}\mathcal{R}_0 = \infty$ and $\lim_{a\rightarrow 1}\mathcal{R}_0 = c^2(1+c^2)^{-1} < 1$.  This, together with Prop. \ref{prop:1}, indicates that when $\sigma = 0$, phytoplankton carrying capacity increases or the zooplankton loss rate decreases, the probability of zooplankton extinction will approach zero.  When phytoplankton carrying capacity is low or the zooplankton loss rate is high, zooplankton extinction is more likely.  

\begin{proposition}
If $\sigma = 0$, then the system \eqref{rescale2} has a unique coexistence equilibrium if and only if $\mathcal{R}_0 > 1$.
\label{prop:2}
\end{proposition}
\begin{proof}
Suppose that for the system \eqref{rescale2}, we have $\sigma = 0$, $$\frac{dp}{d\tau} = \frac{dz}{d\tau} = \frac{dn}{d\tau} = 0,$$ and that $n,p,z > 0.$  Then, the system has the following potential unique coexistence equilibrium:
\begin{equation}
    \E_* = \left\lbrace\begin{split}
       p_* &= \sqrt{\frac{a}{1-a}} \\
        n_* &= (2cs)^{-1}\left[n_b + \sqrt{n_b^2 + 4c^2\tilde{k}\theta s^2}\right] \\
       z_* &= \frac{n_*}{p_*(\tilde{k} + n_*)}\left(1 - \frac{p_*}{c}\right)(1+p_*^2)
    \end{split}\right.
    \label{linCo}
\end{equation}
where $$n_b = -c\tilde{k}s + c\theta s - cp_*\gamma + p_*^2\gamma.$$

\noindent Note that $\mathcal{R}_0 > 1$ if and only if $p_* < c$:
\begin{equation*}
    \begin{split}
        \mathcal{R}_0 = \frac{c^2}{a(1 + c^2)} > 1 &\Leftrightarrow \frac{c^2}{1 + c^2} > a \\ &\Leftrightarrow p_* = \sqrt{\frac{a}{1-a}} < \sqrt{\frac{c^2(1+c^2)^{-1}}{1 - c^2(1+c^2)^{-1}}} = \sqrt{c^2} = c.
    \end{split}
\end{equation*}
Additionally, from its form, it is clear that $n_* > 0$.  Thus, we may conclude that this expression for the coexistence equilibrium is biologically feasible.
\end{proof}

Let
\begin{equation}
\left\lbrace
    \begin{split}
        \psi_1 & = \frac{n_*}{\tilde{k}+n_*}\left(1 - \frac{p_*}{c}\right),~
        \psi_2  = 2p_*\frac{p_*^2}{1+p_*^2}z_*,~ 
        \psi_3  = 2\frac{p_*}{(1+p_*^2)^2}z_*,~\\
        \psi_4  &= \frac{n_*}{\tilde{k}+n_*}\cdot\frac{p_*}{c},
        \psi_5  = \frac{p_*(c-p_*)}{c(\tilde{k} + n_*)}\left(1-\frac{n_*}{\tilde{k}+n_*}\right),~
        \xi_0  = a\tilde{\gamma}\psi_3(s+\gamma \psi_5),\\
        \xi_1 &=  s(-\psi_1 + \psi_3+\psi_4) + \gamma \psi_3\psi_5 + a\tilde{\gamma} \psi_3,~ 
        \xi_2  = -\psi_1 +\psi_3+\psi_4 + \psi_5 + s\\
        \hat{\xi}_1 &= \xi_1 + s\psi_1, \hat{\xi}_2 = \xi_2 + \psi_1,~
        \mathcal{C}_1^1 = \frac{\psi_1 + a\tilde{\gamma} }{\psi_3 + \psi_4 + s\psi_5 + a\tilde{\gamma}},\\ \mathcal{C}_1^2  &= \frac{\xi_0 + \psi_1(\hat{\xi}_1 + s\hat{\xi}_2)}{\hat{\xi}_1\hat{\xi}_2 + s\psi_1^2},~
        \mathcal{C}_1 = \max\{\mathcal{C}_1^1,\mathcal{C}_1^2\}
    \end{split}\right.
    \label{linCoCon}
\end{equation}
where $p_*,n_*,z_*$ are as defined in \eqref{linCo}.
\begin{proposition}
If $\mathcal{R}_0 > 1$ and $\sigma = 0$, then the coexistence equilibrium, $\E_*$ \eqref{linCo}, is locally asymptotically stable if \begin{equation}\mathcal{C}_1:=\max\{\mathcal{C}_1^1,\mathcal{C}_1^2\} < 1 \label{linearCoStab} \end{equation} and has a simple Hopf bifurcation in a parameter of interest, $\alpha$, at value $\alpha_0$  if \begin{equation}\mathcal{C}_1^1 < \mathcal{C}_1^2 = 1 \label{linearHopf1} \end{equation}
and
\begin{equation} -\xi_0'(\alpha_0) + \xi_2(\alpha_0)\xi_1'(\alpha_0) + \xi_1(\alpha_0)\xi_2'(\alpha_0) \neq 0 \label{linearHopf2} \end{equation} 
\label{prop:3}
\end{proposition}

\begin{proof} See appendix A.
% First, note that given system \eqref{rescale} and $\sigma = 0$, the Jacobian matrix evaluated at $\E_*$ is
% \begin{equation}J(\E_*) =
%     \begin{pmatrix}
%     \psi_1 + \psi_2 - \psi_3 - \psi_4 & & & -a & & & \psi_5 \\
%     &  & & & & & \\
%     \tilde{\gamma}z_*(\psi_3 - \psi_2) & & & 0 & & & 0 \\
%     & & & & & & \\
%     -\psi_1 + (1-\gamma)(\psi_3-\psi_2) +  \psi_4 & & & a(1-\gamma) & & & -\left(\psi_5+s\right)
%     \end{pmatrix}
% \end{equation}
% It follows that we have characteristic polynomial (assisted by MatLab Symbolic Math Toolbox) \begin{equation}\chi_{J(\E_*)}(\lambda) = \lambda^3 + \xi_2\lambda^2 + \xi_1\lambda + \xi_0.\end{equation}  Thus we have Hurwitz determinants, $H_i$
% \begin{equation}
%     \begin{split}
%         H_1 & = \xi_2 \\
%         H_2 & = \xi_2\xi_1 - \xi_0 \\
%         H_3 & = \xi_0H_2
%     \end{split}
% \end{equation}

% The generalized Routh-Hurwitz criterion indicates that $\E_*$ will be locally asymptotically stable if and only if $H_i > 0$ $\forall i$.  This is equivalent to condition \eqref{linearCoStab}.  Liu \cite{liu} also indicates that if \begin{equation}H_1 > 0, H_2\vert_{\alpha_0} = 0,\textrm{ and }\frac{d}{d\alpha}H_2\vert_{\alpha_0} \neq 0,\end{equation} then there is a simple Hopf bifurcation in parameter of interest $\alpha$ at point $\alpha_0$.  Thus a necessary condition for a simple Hopf bifurcation to occur is $\xi_0 = \xi_1\xi_2$.  Note condition (4.3.16) is equivalent to conditions \eqref{linearHopf1} and \eqref{linearHopf2}. 
\end{proof}

\begin{figure}[h!]
\centering
    \begin{subfigure}[t]{.3\textwidth}
        
        \includegraphics[width=\textwidth]{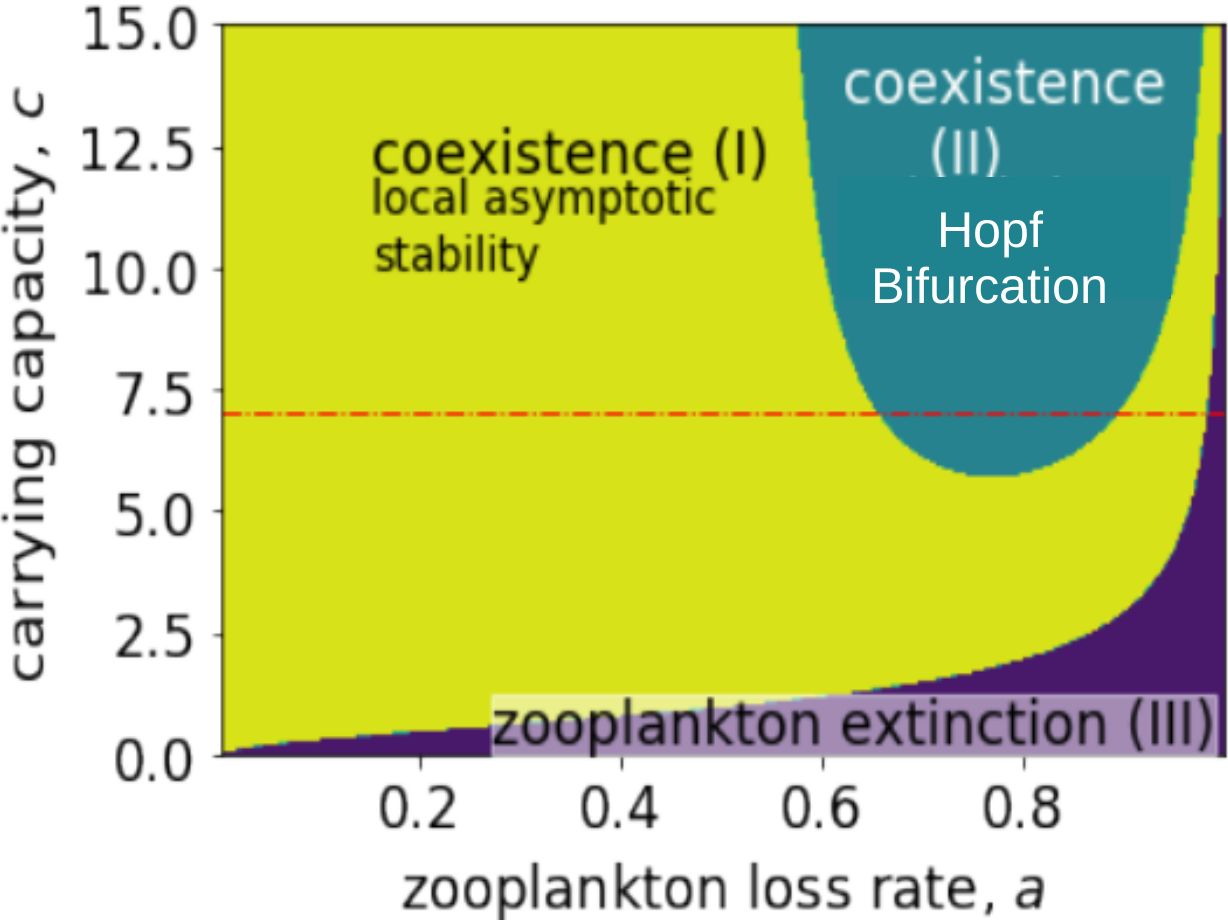}
        \caption{}
    \end{subfigure}
     \begin{subfigure}[t]{.29\textwidth}
        \centering
        \includegraphics[width=\textwidth]{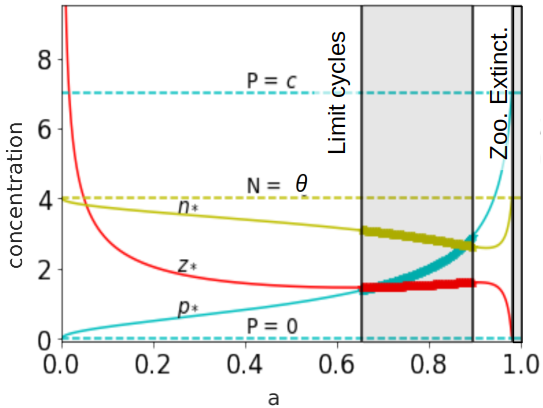}
        \caption{}
    \end{subfigure}
     \begin{subfigure}[t]{.315\textwidth}
        
        \includegraphics[width = \textwidth]{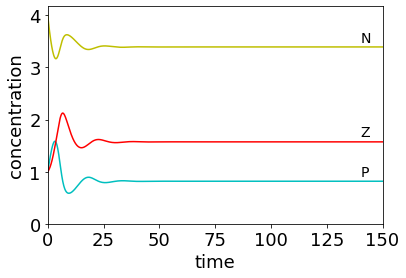}
        \caption{}
    \end{subfigure}
    \begin{subfigure}[t]{.315\textwidth}
        
        \includegraphics[width = \textwidth]{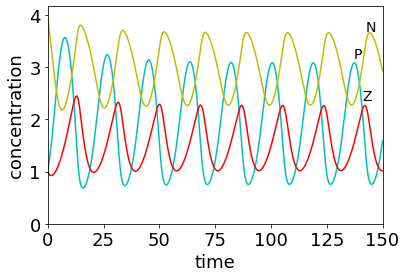}
        \caption{}
    \end{subfigure}
     \begin{subfigure}[t]{.315\textwidth}
                \includegraphics[width = \textwidth]{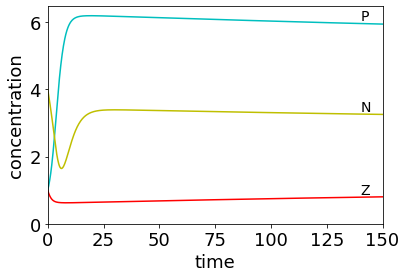}
        \caption{}
    \end{subfigure}
     \begin{subfigure}[t]{.315\textwidth}
        
        \includegraphics[width = \textwidth]{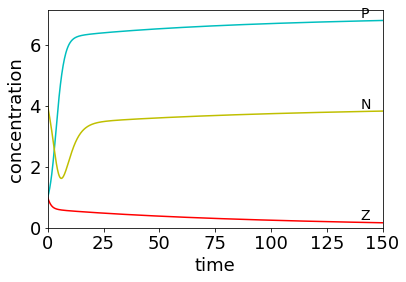}
        \caption{}
    \end{subfigure}
    \caption{\textbf{Bifurcation diagrams and representative simulations for the NPZ system with linear zooplankton mortality}.  Blue, red, and yellow curves represent phytoplankton, zooplankton, and nutrients respectively. \textbf{(a)} Two-dimensional bifurcation diagram with respect to the model parameters $a,c$, representing the zooplankton mortality rate and the phytoplankton carrying capacity, respectively.   \textbf{(b)} One-dimensional bifurcation diagram with respect to the parameter $a$ when $c = 7$ (dashed red line in (a)).  All other model parameters are as specified in table \ref{tab:sims}.  \textbf{(c-f)}  Time-dependent solutions demonstrating the qualitative model dynamics when $c = 7$.  The transitions are region 1 ($a = 0.4$) $\rightarrow$ region 2 ($a = 0.7$) $\rightarrow$ region 1  ($a = 0.98$) $\rightarrow$ region 3 ($a = 0.999$).}
    \label{fig:linear}
\end{figure}

We choose the notation $\mathcal{C}_1$ because, as shown in Prop. 3 (similarly with $\mathcal{C}_2$ in Prop. \ref{prop:5}), it describes the qualitative nature of coexistence in the model with linear zooplankton loss term. While the above expressions (derived from the characteristic polynomial of the Jacobian with MatLab Symbolic Math Toolbox) do not have an easily interpreted biological meaning they are useful in two ways. First, they provide mathematically rigorous bounds for both local asymptotic stability and Hopf bifurcation of the coexistence equilibrium.  Second, these expressions give an indication of which parameters are important to the qualitative dynamics of the model.  Via these expressions, we can see that, in the case of linear zooplankton loss, the only parameters which affect the unique coexistence equilibrium and its asymptotic dynamics are $a$, zooplankton loss rate, and $c$, phytoplankton carrying capacity. 

Hopf bifurcation is defined as the birth of a limit cycle from an equilibrium where the equilibrium changes stability via a pair of purely imaginary eigenvalues. The bifurcation can be supercritical or subcritical, resulting in stable or unstable limit cycles \cite{andronov1971theory}.  In model \eqref{rescale2} with $\sigma = 0$ in the parameter region where Hopf bifurcation occurs we observe that these bifurcations occur around the values of bifurcation parameter $a$ where $p_* \approx z_*$, and act as a transitory state from high to low phytoplankton abundance (see figure \ref{fig:linear}(b)).

\subsection{Linear zooplankton loss term: a special case}
Suppose that nutrient loss/exchange rate $s = 0$, $\tilde{\gamma} = \gamma$, and upon death, zooplankton instantaneously become nutrients.  Then, system \eqref{rescale2} is closed (ie. $\forall \tau \geq 0$, $n(\tau)+p(\tau)+z(\tau) = N_T = p(0)+z(0)+n(0)$).  In this case, we may rewrite the system as:
\begin{equation}
 \left\lbrace   
 \begin{split}
    \frac{dp}{d\tau} &= \frac{N_T-(p+z)}{\tilde{k}+N_T-(p+z)}p\left(1-\frac{p}{c}\right)-\frac{p^2}{1+p^2}z\\
    \frac{dz}{d\tau} &= \gamma\left(\frac{p^2}{1+p^2}-a\right)z\\
    \frac{dn}{d\tau} &= -\frac{N_T-(p+z)}{\tilde{k}+N_T-(p+z)}p\left(1-\frac{p}{c}\right)+(1-\gamma)\frac{p^2}{1+p^2}z +\gamma a z. \\
    \end{split}\right.
    \label{reduction}
\end{equation}
For this system, there are the following boundary equilibria: $\E_0 = (0,0, N_T)^T$, $\E_{N_T} = (N_T,0,0)^T$, and $\E_{c} = (c,0, N_T-c)^T$ (exists if and only if $N_T \geq c$).  Also, note that for this version of the model, the $p=0$ and $z = 0$ planes are invariant and if $n = 0$ (that is, $N_T = p + z$), then \begin{equation}
    \frac{dn}{d\tau} = (1-\gamma)\frac{p^2}{1+p^2}z+\gamma a z \geq 0.
\end{equation}
Hence, the set $B = \{(p,z,n): 0 \leq p+z \leq N_T\}$ is invariant (and indeed $p+z > N_T \Rightarrow n < 0$).
 Thus, we need only consider solutions on this set.  Additionally, because we may write $n(\tau) = N_T - p(\tau) - z(\tau)$, we need only consider the reduced system
 \begin{equation}
 \left\lbrace
     \begin{split}
          \frac{dp}{d\tau} &= \frac{N_T-(p+z)}{\tilde{k}+N_T-(p+z)}p\left(1-\frac{p}{c}\right)-\frac{p^2}{1+p^2}z = f_1(p,z)\\
    \frac{dz}{d\tau} &= \gamma\left(\frac{p^2}{1+p^2}-a\right)z = f_2(p,z) \\
     \end{split}\right.
     \label{2dim}
 \end{equation}
 
 Now, define the zooplankton invasion number:
\begin{equation}
    \mathcal{R}_1 := \frac{1}{a}\cdot\frac{N_T^2}{1+N_T^2}
\end{equation}
\begin{proposition}
If $N_T < c$ and $\mathcal{R}_1 < 1$, then the zooplankton extinction equilibrium $\E_{N_T}$ is globally asymptotically stable in $B\backslash\{p = 0\}$.  If $N_T > c$ or $\mathcal{R}_1 > 1$, it is unstable.   \label{GAS}
\end{proposition}
\begin{proof}
First, note that the lines $p = 0$ and $z = 0$ are invariant.  Then observe that the Jacobian of system \eqref{2dim} evaluated at $\E_{N_T}$ is
\begin{equation}
    J(\E_{N_T}) = \begin{pmatrix} 
\frac{-N_T(c - N_T)}{c\tilde{k}} & * \\
0 & \gamma a (R_1 - 1) 
\end{pmatrix},
\end{equation}
from this it is clear that if $N_T < c$ and $\mathcal{R}_1 < 1$, then $\mathcal{E}_{N_T}$ is locally stable, and is unstable if either $N_T > c$ or $\mathcal{R}_1 > 1$.

Now observe that for this version of the model, the community collapse equilibrium $\E_0$ is always unstable since
\begin{equation}
J(\E_0) =\begin{pmatrix} 
 \frac{N_T}{\tilde{k}+N_T} & 0 \\
0 & -a\gamma
\end{pmatrix}.
\end{equation}
Note that 
\begin{equation}
\begin{split}
           \mathcal{R}_1 = \frac{N_T^2}{a(1 + N_T^2)} < 1 &\Leftrightarrow \frac{N_T^2}{1 + N_T^2} < a \\ &\Leftrightarrow p_* = \sqrt{\frac{a}{1-a}} > \sqrt{\frac{N_T^2(1+N_T^2)^{-1}}{1 - N_T^2(1+N_T^2)^{-1}}} = \sqrt{N_T^2} = N_T,
    \end{split}
\end{equation}
where $p_*$ is the $p$ component of any possible coexistence equilibrium.
Thus, when $N_T < c$ and $\mathcal{R}_1 < 1$, there are only two equilibria in $B$: $\E_0$ and $\E_{N_T}$.  Since the model is closed, all of our solutions are bounded for $\tau \geq 0$. Thus, any solution contains $[0,\infty)$ in its domain and has a compact and non-empty $\omega$-limit set.

Note that for
$\phi = 1/p^2$,  $N_T < c$, and $\mathcal{R}_1 < 1$, for any solution to \eqref{reduction} $(p,z)^T \in~    B\backslash(\{p =0\}\cup\{z = 0\})$:
\begin{equation}
\begin{split}
    \frac{\partial}{\partial p}\left[\phi\frac{dp}{d\tau}\right] &= -\frac{1-p/c}{p}\left(\frac{\tilde{k}}{(\tilde{k}+N_T-(p+z))^2}\right) - \frac{N_T-(p+z)}{\tilde{k}+N_T-(p+z)}\cdot\frac{1}{p^2}\\ &\hspace{1.5in}-\frac{2p}{(1+p^2)^2}z < 0 \\
    \frac{\partial}{\partial z}\left[\phi\frac{dz}{d\tau}\right] &= \gamma\left(\frac{p^2}{1+p^2}-a\right)\frac{1}{p^2} \leq \frac{\gamma a}{p^2}\left(\mathcal{R}_1-1\right) < 0
    \end{split}
\end{equation}
since $\frac{p^2}{1+p^2}$ is an increasing function of $p$.  Hence, when  $N_T < c$ and $\mathcal{R}_1 < 1$,
\begin{equation}
    \nabla\cdot(\phi(p)f(p,z)) = \frac{\partial}{\partial p}\left[\phi\frac{dp}{d\tau}\right] + \frac{\partial}{\partial z}\left[\phi\frac{dz}{d\tau}\right] < 0 \hspace{.1in} \forall (p,z)^T \in B\backslash(\{p =0\}\cup\{z = 0\}).
\end{equation}
Thus, by Dulac's criterion, it follows that there are no closed orbits wholly contained in $B\backslash(\{p =0\}\cup\{z = 0\})$.

Now, notice that in the invariant line $N_T = n+p$ (ie. $z = 0$), $\frac{dz}{d\tau} = 0$ and if $p > 0$, then $\frac{dp}{d\tau} > 0$ when $p < N_T < c$.  Similarly, on the invariant line $N_T = n+z$ ($p = 0)$, $\mathcal{E}_0$ attracts all solutions.  Let $L$ stand for the $\omega$-limit set of some point $(p_0,z_0,n_0)^T\in B\backslash(\{p =0\}\cup\{z = 0\})$. Recall that there are two equilibia, $\E_{0}$ and $\E_{N_T}$, in $B$. Because $\E_{0}$ is a hyperbolic saddle-node it cannot belong to any heteroclinic cycle or homoclinic loop.  Consequently, there are no heteroclinic cycles or homoclinic loops at all (Poincare–Bendixson theorem).  Hence, $L = \{\E_{N_T}\}$.  It follows that $\E_{N_T}$ is globally asymptotically stable in $B\backslash\{p = 0\}$.  \\
\end{proof}

\begin{proposition}
  The zooplankton extinction equilibrium $\E_c = (c,0,N_T - c)^T$ exists if and only if $N_T \geq c$,.  Further, if $\mathcal{R}_0 < 1$ and $N_T > c$, then it is locally asymptotically stable.  If $\mathcal{R}_0 > 1$, it is unstable.  
  \label{prop:special}
\end{proposition}
\begin{proof} 
See appendix A.

\end{proof}

\subsection{Quadratic zooplankton loss ($\sigma = 1$)} 
\label{sec:3b}
Suppose that $n,p,z > 0$ and $$\frac{dp}{d\tau} = \frac{dz}{d\tau} = \frac{dn}{d\tau} = 0. $$  It follows that in the case of quadratic zooplankton loss, any possible coexistence equilibrium must satisfy

\begin{equation}
  \left\lbrace \begin{split}
       p_* &= \sqrt{\frac{az_*}{1-az*}},  \\ 
       n_* &=  s^{-1}(s\theta - \gamma a z_*^2),    \\
       0 &= az_*^2 + z_* +\dfrac{n_*}{k+n_*}p_* \left(1-\frac{p_*}{c}\right),\end{split}\right.
       \label{quadCo}
\end{equation}
which cannot be solved explicitly.  
Because of this, we now make use of the following result from general persistence theory:\\

\noindent\textbf{Theorem} (Existence of coexistence equilibrium, \cite{hsmith}) Suppose that 
\begin{enumerate}
    \item $X$ is a closed, convex subset of a Banach Space,
    \item $\phi$ has a compact attractor, $B$, of bounded subsets in $X$,
    \item $\rho$ is continuous and concave,
    \item $\phi$ is uniformly weakly persistent,
    \item $\phi(t,\cdot)$ is compact for some $t > 0$.
\end{enumerate}  Then, there exists an equilibrium $x^*$ with $\rho(x^*) > 0$.  \\

\begin{proposition}
There exists at least one coexistence equilibrium in the system \eqref{rescale2} when $\sigma = 1$.
\label{prop:4}
\end{proposition}
\begin{proof}
We first show that the system \eqref{rescale2} is dissipative.
Note that the planes where $p = 0$ and $z = 0$ are invariant, and when $n = 0$, $\frac{dn}{d\tau} > 0$. Then, define
\begin{equation}
        A_p = \{(p,z,n)^T: p \geq c, z> 0, n > 0\} 
    \label{Ap}
\end{equation}
and note that $\forall (p,z,n)^T \in A_p$, $\frac{dp}{d\tau} < 0$. Now define \begin{equation}
    \begin{split}
        A_z^{(2)} &= \{(p,z,n)^T: 0 < p\leq c, z > \mathcal{R}_0, n > 0\} \\
        A_n &= \{(p,z,n)^T: 0 < p \leq c, 0 < z \leq \mathcal{R}_0, n > \hat{n}\}
    \end{split}
\end{equation}
where $\hat{n} = s^{-1}((1-\gamma)a\mathcal{R}_0^2 + s\theta)$.
First, note that $\forall$ $(p,z,n)^T \in A_z^{(2)}$ \begin{equation}  \frac{dz}{d\tau} \leq \tilde{\gamma}a(\mathcal{R}_0 - z)z < 0.\end{equation}  Next, observe that $\forall (p,z,n)^T \in A_n$,
\begin{equation}
\begin{split}
    \frac{dn}{d\tau} &\leq (1-\gamma)\frac{c^2}{1+c^2}\mathcal{R}_0 + s(\theta - n) \\
    &= (1-\gamma)a\mathcal{R}_0^2 + s(\theta - n) \\
    &< 0.
    \end{split}
\end{equation}
Hence $\forall (p,z,n)^T$ such that $p_0,z_0,n_0 \geq 0$, $$\lim_{\tau\rightarrow\infty}(p(\tau),z(\tau),n(\tau))^T \in B = \{(p,z,n): 0 \leq p \leq c, 0 \leq z \leq \mathcal{R}_0, 0 \leq n \leq \hat{n}\}. $$  Thus, system \eqref{rescale2} is dissipative.

Note that when $\sigma = 1$, the Jacobian evaluated at equilibrium $\E_{np}$ takes the form 

  \begin{equation}
    J(\E_{np}) = 
    \begin{pmatrix} 
    A & 0 \\ * & - s 
    \end{pmatrix},
\end{equation}
where $$A = 
\begin{pmatrix}
-\frac{\theta}{k+\theta} & -a\mathcal{R}_0 \\
& \\
0 & \tilde{\gamma}a\mathcal{R}_0 
\end{pmatrix}.$$
Thus if $\sigma = 1$, $\E_{np}$ is always a hyperbolic saddle node.

Next we show that the system is robustly uniformly $\rho$-persistent, where $\rho = \min\{n(\tau),p(\tau),z(\tau)\}$.

We may consider the equilibria of the system as trivial periodic solutions.  Applying corollary 4.7 of \cite{salceanu}, we see, via proposition 4.1 and theorem 3.2 of \cite{salceanu}, that phytoplankton are robustly persistent if the eigenvalue in the $p$ direction of $J(\E_n)$, $\lambda_p^{\E_{n}} > 0$.  Similarly, since zooplankton depend upon phytoplankton as a resource, we need only show the eigenvalue in the $z$ direction of $J(\E_{np})$, $\lambda_z^{\E_{np}} > 0$ provided the first condition holds.  Note that 	$$J(\E_n) = \begin{pmatrix}
		\frac{\theta}{\tilde{k} + \theta} & 0 & 0\\
		\\
		0 & -\tilde{\gamma}a(1-\sigma) & 0 \\ \\
		-\frac{\theta}{\tilde{k} + \theta} & 0 & -s
	\end{pmatrix}. $$
	From this it is clear that $\lambda_p^{\E_{n}} = \theta(\tilde{k} + \theta)^{-1} > 0$.
	Next, notice that boundary equilibrium $\E_{np}$ has  $\lambda_z^{\E_{np}} = \eta\gamma\beta^{-1}a(\mathcal{R}_0 - (1-\sigma))$.  From this, it is clear that this eigenvalue is strictly positive if $\sigma = 1$ or $\mathcal{R}_0 > 1$ and $\sigma = 0$. 

 \noindent Hence, all of the conditions for the existence of a coexistence equilibrium are either met or exceeded.
\end{proof}
% Careful consideration of coexistence equilibrium constraint \eqref{quadCo} and generation of bifurcation diagrams (see GitHub \cite{MacdonaldCode}) indicate that $\theta > 0$ and the quadratic loss are what allow for the possibility of the existence of multiple coexistence equilibria (see bifurcation diagrams in Fig. \ref{fig:quadratic}).
% \begin{proposition}
% If $\sigma = 1$ then the community collapse, equilibrium, $\E_n$, is always unstable 
% \end{proposition}
 
% \begin{proof}
% See appendix.
% \end{proof}

% \begin{proposition}
% When $\sigma = 1$ there always exists at least one coexistence equilibrium.
% \end{proposition}

% \begin{proof}
% Recall that our system is well-posed and dissipative (Theorem 1).  Clearly, our choice of $\rho$ is continuous and concave (See appendix.).  Finally, by Theorem 4 it follows that the system is uniformly weakly persistent and $\phi$ is compact from some $t > 0.$  The desired result follows.
% \end{proof}

% The coexistence equilibria of the system (3) with quadratic loss are found by solving the following system of equations:  
% \begin{equation}
%   \left\lbrace \begin{split}
%       z_* &= \frac{p_*^2}{a(1+p_*^2)}  \\ \\
%       n_* &=  \frac{1}{s}\left(sn_i - \gamma\frac{p_*^2}{1+p_*^2}z_*\right) \\ \\
%         \frac{p_*^3}{a(1+p_*^2)^2} &= \frac{n_*}{(\tilde{k} + n_*)}\left(1 - \frac{p_*}{c}\right)
%   \end{split} \right. 
% \end{equation}It is not possible to find an explicit expression for each $p_*$.  

For arbitrary coexistence equilibrium $\E_* = (p_*,z_*n_*)^T$ satisfying \eqref{quadCo}, let
\begin{equation}
\left\lbrace
    \begin{split}
         \nu_0 & = a\tilde{\gamma}z_*\left[(\psi_3+\psi_4)(\psi_5 + s)+\psi_5(\psi_1 + \psi_2(1-\gamma))+\psi_2\psi_5\gamma z_* + \psi_3sz_*\right]\\
         &\hspace{.3in}- a\tilde{\gamma}z_*\left[(\psi_1+\psi_2)(\psi_5+s) + \psi_5(\psi_3(1-\gamma) + \psi_4) + \psi_3\psi_5\gamma z_* + \psi_2sz_*\right] \\
        \nu_1 & =  (\psi_3 + \psi_4)(a\tilde{\gamma}z_*+\psi_5+s) + a\tilde{\gamma}z_*(\psi_5 + s) + \psi_1\psi_5 + \psi_2\psi_5(1-\gamma) \\ &\hspace{.3in}-\left((\psi_1+\psi_2)(a\tilde{\gamma}z_*+\psi_5+s) + a\tilde{\gamma}z_*^2\psi_2 +\psi_3(1-\gamma)+\psi_4 \right)\\
        \nu_2 & =  \psi_3 + \psi_4 + a\tilde{\gamma}z_* + s -(\psi_1 + \psi_2)\\
        \hat{\nu}_0 &= \nu_0 + a\tilde{\gamma}z_*\left[(\psi_1+\psi_2)(\psi_5+s) + \psi_5(\psi_3(1-\gamma) + \psi_4) + \psi_3\psi_5\gamma z_* + \psi_2sz_*\right], \\ \hat{\nu}_1 &= \nu_1 + \left((\psi_1+\psi_2)(a\tilde{\gamma}z_*+\psi_5+s) + a\tilde{\gamma}z_*^2\psi_2 +\psi_3(1-\gamma)+\psi_4 \right), \\
        \hat{\nu}_2 &= \nu_2 +  \psi_1 + \psi_2,~
        \mathcal{C}_2^1 = \frac{\hat{\nu}_2 - \nu_2 }{\hat{\nu}_2},~
        \mathcal{C}_2^2 = \frac{\hat{\nu}_0 + \hat{\nu}_1(\hat{\nu}_2 - 
        \nu_2)+ \hat{\nu}_2(\hat{\nu}_1 - \nu_1)}{\hat{\nu}_1\hat{\nu}_2 + (\hat{\nu}_1-\nu_1)(\hat{\nu}_2-\nu_2) + (\hat{\nu}_0 - \nu_0)},~ \\
        \mathcal{C}_2^3  &= \frac{\hat{\nu}_0 - \nu_0 }{\hat{\nu}_0},~
        \mathcal{C}_2 = \max\{\mathcal{C}_2^1,\mathcal{C}_2^2,\mathcal{C}_2^3\},
    \end{split}\right.
\end{equation}
where $\psi_i$ are as defined in system \eqref{linCoCon} and expressions were again derived from the characteristic polynomial of the Jacobian with MatLab Symbolic Math Toolbox.

\begin{proposition}
$\E_*$ is locally asymptotically stable if \begin{equation}\mathcal{C}_2 < 1 \end{equation} and has a simple Hopf bifurcation in a parameter of interest, $\alpha$ at point $\alpha_0$,  if \begin{equation}\mathcal{C}_2^1 < \mathcal{C}_2^2 = 1 \label{HopfQuad}\end{equation}
and
\begin{equation} -\nu_0'(\alpha_0) + \nu_2(\alpha_0)\nu_1'(\alpha_0) + \nu_1(\alpha_0)\nu_2'(\alpha_0) \neq 0\end{equation} 
\label{prop:5}
\end{proposition}

\begin{proof} 
See appendix A.
\end{proof}
As with Prop. \ref{prop:3}, while it is difficult to directly interpret these quantities biologically, they indicate that many more parameters can affect the qualitative dynamics of the model.

   \begin{table}[h!]
    \centering
    \caption{Summary of threshold analysis from Sec. \ref{sec:3}}
    \begin{tabular}{ll}
\hline\noalign{\smallskip}
       Equilibrium $(p,z,n)^T$& Description  \\
 \noalign{\smallskip}\hline\noalign{\smallskip}
        $\mathcal{E}_{np} = (c,0,\theta)^T$ & Boundary (zooplankton extinction),  LAS if $\mathcal{R}_0 < 1$ and $\sigma = 0$ \\ &  unstable  otherwise  (Props. \ref{prop:1}, \ref{prop:4})\\ & \\
        $\mathcal{E}_{n} = (0,0,\theta)^T$ & Boundary (community collapse), unstable  (Prop. \ref{prop:append})\\ & \\
        $\E_{N_T} = (N_T,0,0)^T$ & Special case, boundary (zooplankton extinction) \\ & GAS if $\mathcal{R}_1 < 1$ and $N_T < c$ (Prop. \ref{GAS}). \\
        & \\
        $\E_{c} = (c,0,N_T-c)^T$ & Special case, boundary (zooplankton extinction) \\ & exists and LAS if $\mathcal{R}_0 < 1$ and $N_T > c$.  \\ & (Prop. \ref{prop:special}). \\
        & \\
        $\mathcal{E}_* = (p_*,z_*,n_*)^T$ & Coexistence ($\sigma = 0)$: \\ & Unique existence iff $\mathcal{R}_0 > 1$  (Prop. \ref{prop:2})\\ 
        & LAS if $\mathcal{C}_1 < 1$ (Prop. \ref{prop:3})\\
        & Simple Hopf bifurcation if $\mathcal{C}_1^1 < \mathcal{C}_1^2 = 1$, condition \eqref{linearHopf2}  is satisfied\\
        &(Prop. \ref{prop:3}) \\
        & \\

       & Coexistence ($\sigma = 1)$: \\
        & Exists at least one (Prop. \ref{prop:4})\\
        & For arbitrary coexistence equilibrium: \\
        &LAS if $\mathcal{C}_2 < 1$ (Prop. \ref{prop:5})\\
        & Simple Hopf bifurcation if $\mathcal{C}_2^1 < \mathcal{C}_2^2 = 1$, condition \eqref{HopfQuad} is satisfied \\ & (Prop. \ref{prop:5})\\
    \hline\noalign{\smallskip}
   \end{tabular}
    \label{tab:threshold}
\end{table}
\begin{figure}[h!]
    \centering
       \begin{subfigure}[t]{\textwidth}
       \centering
        \includegraphics[width=.6\textwidth]{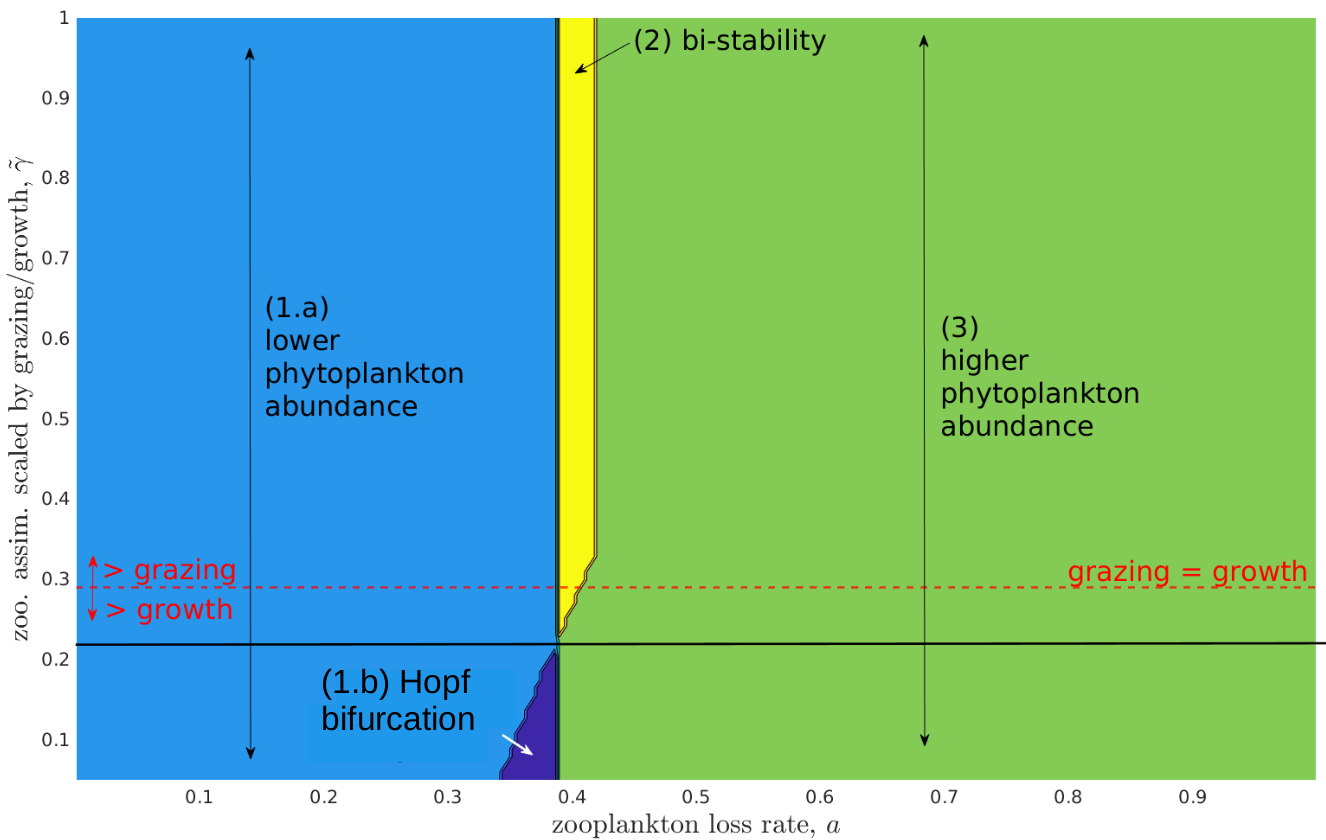}
        \caption{}
    \end{subfigure}
      \begin{subfigure}[t]{.3\textwidth}
        
        \includegraphics[width = \textwidth]{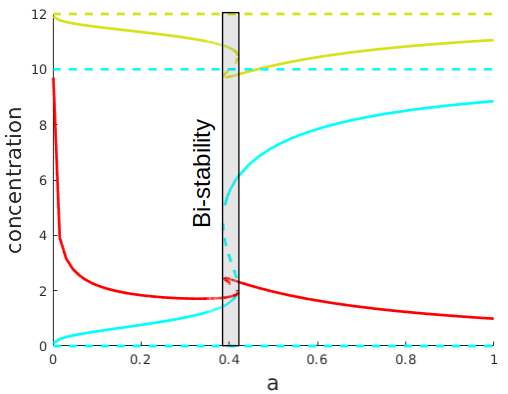}
        \caption{}
    \end{subfigure}
     \begin{subfigure}[t]{.3\textwidth}
        \centering
        \includegraphics[width=\textwidth]{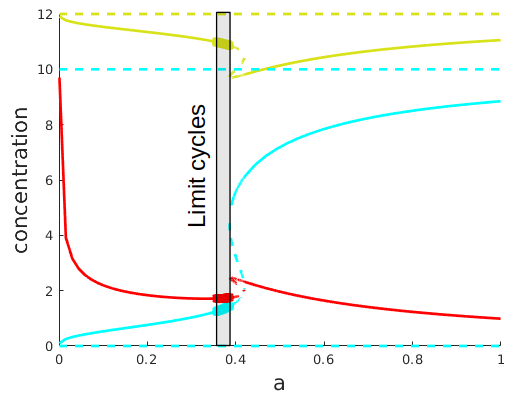}
        \caption{}
    \end{subfigure}
    \begin{subfigure}[t]{.32\textwidth}
        
        \includegraphics[width = \textwidth]{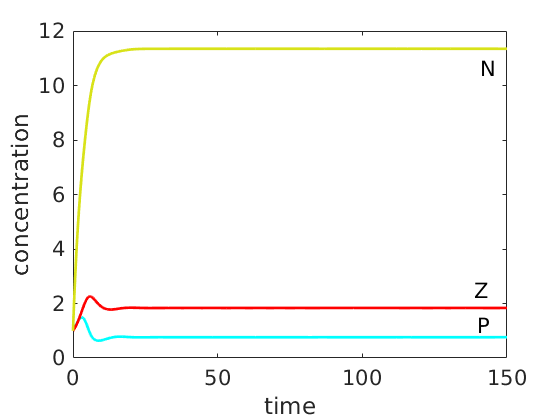}
        \caption{}
    \end{subfigure}
         \begin{subfigure}[t]{.32\textwidth}
        
        \includegraphics[width = \textwidth]{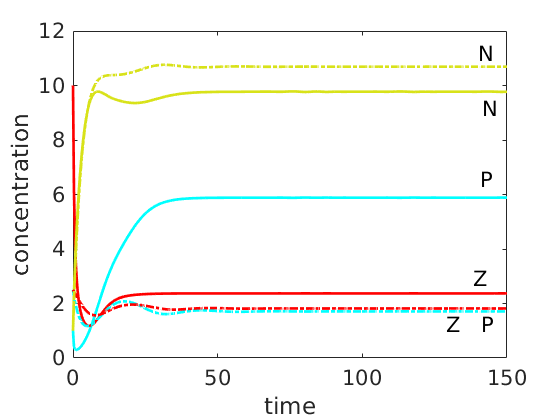}
        \caption{}
    \end{subfigure}
        \begin{subfigure}[t]{.32\textwidth}
        
        \includegraphics[width = \textwidth]{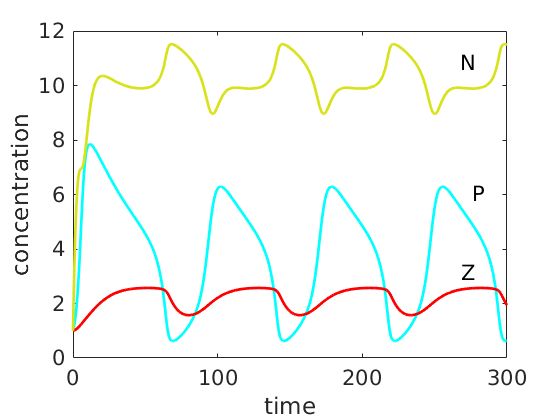}
        \caption{}
    \end{subfigure}
        \begin{subfigure}[t]{.32\textwidth}
        
        \includegraphics[width = \textwidth]{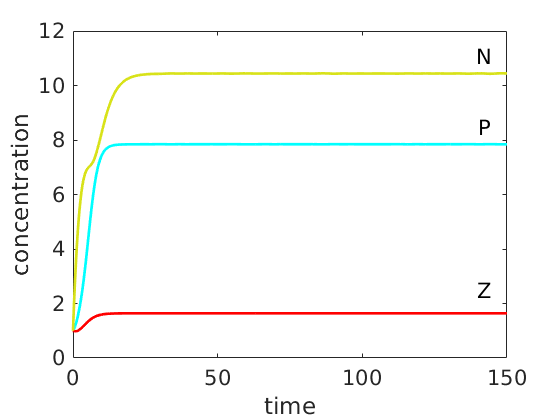}
        \caption{}
    \end{subfigure}
     \caption{\textbf{Bifurcation diagrams for the NPZ subsystem with quadratic zooplankton mortality.}  Blue, red, and yellow curves represent phytoplankton, zooplankton, and nutrients respectively. \textbf{(a)} Two-dimensional bifurcation diagram with respect to the model parameters $\tilde{\gamma}$, and  $a$. One-dimensional bifurcation diagrams in $a$, where the transitory state between low and high phytoplankton abundance is either \textbf{(b)} bi-stability with $\tilde{\gamma}$ = 0.29, or \textbf{(c)} Hopf bifurcation with $\tilde{\gamma}$ = 0.1.  All other parameter values are given in table \ref{tab:sims}. \textbf{(d-g)} Time-depentent solutions of the qualitative model dynamics in different parameter regions. The transition from low to high phytoplankton abundance is either region 1a ($ a = 0.3, \tilde{\gamma} = 0.29$) $\rightarrow$  region 2 ($ a = 0.41, \tilde{\gamma} = 0.29$) $\rightarrow$ region 3 ($ a = 0.6, \tilde{\gamma} = 0.29$) or region 1a  $\rightarrow$ region 1b ($ a = 0.37, \tilde{\gamma} = 0.1$) $\rightarrow$ region 3 dependent on the value of $\tilde{\gamma}$.}
     \label{fig:quadratic}
\end{figure}
In the case of a quadratic zooplankton loss term, studying these equations indicates a much broader subset of the model parameters can affect both the number of coexistence equilibria and the asymptotic dynamics of each such equilibrium.  Specifically from the numerical generation of bifurcation diagrams (see GitHub \cite{MacdonaldCode}), we can see that the parameters which can affect both stability and number of equilibria are again $a$ and $c$ as well as the nutrient loss/exchange rate, $s$; the intrinsic nutrient level, $\theta$; the zooplankton assimilation rate, $\gamma$; and the nutrient update half saturation constant, $\tilde{k}$.  In addition $\tilde{\gamma}$, which $\gamma$ scaled by the ratio of maximum zooplankton grazing rate, $\eta$, to the phytoplankton response to light, $\beta$, can affect the asymptotic behavior of the system.

\section{Extending threshold analysis to the NPZD Model and ecological implications}
\label{sec:eco_imp}
Here, we investigate the analytical and numerical properties of the full NPZD system \eqref{rescale} along with ecological implications of phytoplankton overpopulation during HABs and the effects of ecological disturbances on these dynamics. 

Define phytoplankton persistence number
\begin{equation}
    \mathcal{P}_0^p := \frac{1}{\tilde{\epsilon}}\cdot\frac{\theta}{\tilde{k}+\theta}
\end{equation}

\begin{proposition}
If $\mathcal{P}_0^p > 1$ then for the model \eqref{rescale} the phytoplankton population is robustly uniformly $\rho-$persistent where $\rho = \min_\tau p(\tau)$.
\label{prop:8}
\end{proposition}

\begin{proof}
See appendix A.
\end{proof}

We see that for this model $\mathcal{P}_0^p$ is the average lifespan of phytoplankton scaled by a ratio of the intrinsic nutrient level of the system, with persistence being assured if $\tilde{\epsilon} < \theta/(k+\theta)$.  We particularly focus on its implications for the zooplankton extinction equilibrium as well as the conditions for zooplankton persistence and the existence of at least one coexistence equilibrium.

The model \eqref{rescale} has two boundary equilibria, zooplankton extinction equilibrium $\mathcal{E}_{npd} = \left(\hat{p},0,\theta,\hat{d}\right)^T$, where

\begin{equation}
  \left\lbrace  \begin{split}
        \hat{p} &= c\left(1-\frac{1}{\mathcal{P}_0^p}\right) \\
        \hat{d} &= \frac{\tilde{\epsilon} c}{\psi}\left(1-\frac{1}{\mathcal{P}_0^p}\right), \\
\end{split}\right.
\end{equation}

and community collapse equilibrium $\mathcal{E}_n' = (0,0,\theta,0)^T$.  Now define the zooplankton invasion number
\begin{equation}
    \mathcal{R}_0^z := \max\{\mathcal{R}_{0,1}^z,\mathcal{R}_{0,2}^z,\mathcal{R}_{0,3}^z,\mathcal{R}_{0,4}^z\},
\end{equation}
where
\begin{equation}
\left\lbrace\begin{split}
     b_1 &= \tilde{\alpha} c(1-1/\mathcal{P}_0^p), b_2 = \frac{c\tilde{k}(\mathcal{P}_0^p-1)}{(\mathcal{P}_0^p(\tilde{k}+\theta))^2},
        b_3 = \frac{c(1-1/\mathcal{P}_0^p)}{\mathcal{P}_0^p(\tilde{k}+\theta)}\left(1-\tilde{\epsilon}\right)+s \\
\mathcal{R}_{0,1}^z &:= \frac{c(1-1/\mathcal{P}_0^p)(\xi+c(1-1/\mathcal{P}_0^p))}{\tilde{\alpha}(1+c^2(1-1/\mathcal{P}_0^p)^2)}, 
    \mathcal{R}_{0,2}^z := \frac{\tilde{\epsilon}(\mathcal{P}_0^p - 1)}{b_3 + \psi} \\
    \mathcal{R}_{0,3}^z &:= \frac{((b_3+\psi)(b_2+b_3+\psi) + \tilde{\epsilon}(\tilde{\epsilon}b_2 + \psi b_3))(\mathcal{P}_0^p-1)}{(b_3 + \psi)(\tilde{\epsilon}b_2+\psi b_3) + \tilde{\epsilon}((\mathcal{P}_0^p-1)(b_2+b_3+\psi) + \psi(b_2+b_3))(\mathcal{P}_0^p-1)} \\
    \mathcal{R}_{0,4}^z &:= \frac{\mathcal{R}_{0,3}^z + \mathcal{R}_0^{p}}{\mathcal{R}_{0,3}^z\mathcal{R}_{0}^p+ 1} 
\end{split}\right.
\end{equation}  These expressions are complex and so difficult to interpret directly, though we observe that they are each functions of $\mathcal{P}_0^p$ (as well as other model parameters).
% \begin{equation}
% \left\lbrace\begin{split}
%      b_1 &= \alpha c(1-1/\mathcal{P}_0^p),
%         b_2 = \frac{c\tilde{k}(\mathcal{P}_0^p-1)}{(\mathcal{P}_0^p(\tilde{k}+\theta))^2}, 
%         b_3 = \frac{c(1-1/\mathcal{P}_0^p)}{\mathcal{P}_0^p(\tilde{k}+\theta)}\left(1-\epsilon\right)+s \\
% \mathcal{R}_{0,1}^z &:= \frac{c(1-1/\mathcal{P}_0^p)(\xi+c(1-1/\mathcal{P}_0^p))}{\alpha(1+c^2(1-1/\mathcal{P}_0^p)^2)},
%     \mathcal{R}_{0,2}^z := \frac{\tilde{\epsilon}(\mathcal{P}_0^p - 1)}{b_3 + \psi} \\
%     \mathcal{R}_{0,3}^z &:= \frac{((b_3+\psi)(b_2+b_3+\psi) + \tilde{\epsilon}(\tilde{\epsilon}b_2 + \psi b_3))(\mathcal{P}_0^p-1)}{(b_3 + \psi)(\tilde{\epsilon}b_2+\psi b_3) + \tilde{\epsilon}((\mathcal{P}_0^p-1)(b_2+b_3+\psi) + \psi(b_2+b_3))(\mathcal{P}_0^p-1)} \\
%     \mathcal{R}_{0,4}^z &:= \frac{\mathcal{R}_{0,3}^z + \mathcal{P}_0^{p}}{\mathcal{R}_{0,3}^z\mathcal{P}_{0}^p+ 1},
% \end{split}\right.
% \end{equation}

% \begin{equation}
%     \left\lbrace
%     \begin{split}
%          z_* &=\frac{(1-\alpha)p_*^3 + \xi p_*^2 - \alpha p_*}{a(1+p_*^2)(\xi+p_*)}\\
%       d_* &= \frac{1}{\psi}\left[(1-\gamma)\frac{p_*^2}{1+p_*^2}z_* + \tilde{\epsilon}p_*\right] \\
%       n_* &= \frac{1}{2cs}\left[n_b+\sqrt{n_b^2+4c^2s\tilde{k}(s\theta +\psi d_*)}\right]
%     \end{split}
%     \right.
% \end{equation}

% where $$n_b = -(p(c-p) + c(s(\theta-k) + \psi d_*)) $$

% $$\left[1-(az_*^\sigma +\tilde{\alpha})\right]p_*^3 + \xi\left[1-az_*^\sigma\right]p_*^2 - [az_*^\sigma+\tilde{\alpha}]p_* - az_*^\sigma\xi$$

\begin{proposition}
For the model \eqref{rescale}, boundary equilibrium $\mathcal{E}_{npd}$ exists if and only if $\mathcal{P}_0^p > 1$.  It is locally asympotically stable if $\mathcal{R}_0^z < 1$, and is unstable if $\mathcal{R}_0^z > 1$.
\end{proposition}
\begin{proof}
The existence of $\mathcal{E}_{npd}$ if and only if $\mathcal{P}_0^p > 1$ is clear from its form.  Note that the Jacobian of model \eqref{rescale} evaluated at $\mathcal{E}_{npd}$ is
\begin{equation}J(\mathcal{E}_{npd}) = \begin{pmatrix}
\tilde{\epsilon}\left(\mathcal{P}_0^p -1\right) & -\frac{\tilde{\alpha} c(1-1/\mathcal{P}_0^p)}{\xi + c(1-1/\mathcal{P}_0^p)}\mathcal{R}_{0,1}^z & \frac{c\tilde{k}(\mathcal{P}_0^p-1)}{(\mathcal{P}_0^p)^2(\tilde{k}+\theta)^2} & 0 \\
0 & \tilde{\alpha}\tilde{\gamma}c(1-1/\mathcal{P}_0^p)(\mathcal{R}_{0,1}^z-1) & 0 & 0 \\
\tilde{\epsilon}(\mathcal{P}_0^p - 2) & 0 & -\left(\frac{c(1-1/\mathcal{P}_0^p)}{\mathcal{P}_0^p(\tilde{k}+\theta)}\left(1-\tilde{\epsilon}\right)+s\right) & \psi\\
\tilde{\epsilon} & (1-\gamma)\frac{\tilde{\alpha} c(1-1/\mathcal{P}_0^p)}{\xi + c(1-1/\mathcal{P}_0^p)}\mathcal{R}_{0,1}^z & 0 & -\psi 
\label{npdJac}
\end{pmatrix}.\end{equation}
which has characteristic polynomial 
\begin{equation}
\begin{split}
    \chi_{J(\E_{npd})}(\lambda) &= 
    f(\lambda)\cdot g(\lambda) \\ &=(\lambda - b_1(\mathcal{R}_{0,1}^z-1))[\lambda^3 + (b_3 + \psi -\tilde{\epsilon}(\mathcal{P}_0^p - 1))\lambda^2  \\
    &\vspace{1in}(\tilde{\epsilon}(b_2 - (\mathcal{P}_0^p-1)(b_2+b_3+\psi)) + \psi b_3)\lambda-\tilde{\epsilon}\psi(b_2+ b_3)(\mathcal{P}_0^p - 1)].
\end{split}
\end{equation}  

The positivity of $b_1,b_2,b_3$ is assured by $\mathcal{P}_0^p > 1$ and the form of $\mathcal{P}_0^p$. The sign of the eigenvalue $\lambda_1 = b_1(\mathcal{R}_{0,1}^z-1)$ is clear.  Thus we turn to Hurwitz determinants to find conditions on the sign of the real parts of the roots of $g(\lambda)$:
\begin{equation}
    \begin{split}
        H_1 &= b_3 + \psi - \tilde{\epsilon}(\mathcal{P}_0^p-1) \\
        H_2 &= H_1(\tilde{\epsilon}(b_2 - (\mathcal{P}_0^p-1)(b_2+b_3+\psi)) + \psi b_3) + \tilde{\epsilon}\psi(b_2+b_3)(\mathcal{P}_0^p-1) \\
        H_3 &= -\tilde{\epsilon}\psi(b_2+ b_3)(\mathcal{P}_0^p - 1)H_2
    \end{split}
\end{equation}
Next note that for Hurwitz determinant $H_i$, $1 \leq i \leq 3$, $H_i > 0$ if and only if $R_{0,i+1} < 1$.  The desired result follows.
\end{proof}

Now define zooplankton persistence number 
\begin{equation}
  \mathcal{P}_0^z := \min\{\mathcal{P}_0^p,\mathcal{R}_{0,1}^z\},
\end{equation}

where \begin{equation} 
\begin{split}
\mathcal{R}_{0,1}^z &=  \frac{c(1-1/\mathcal{P}_0^p)(\xi+c(1-1/\mathcal{P}_0^p))}{\tilde{\alpha}(1+c^2(1-1/\mathcal{P}_0^p)^2)} \\
 &= \frac{\hat{p}(\xi+\hat{p})}{\tilde{\alpha}(1 + \hat{p}^2)} \\
\end{split}
\end{equation}

Much like the quadratic zooplankton loss term case of model \eqref{rescale2}, an explicit expression for the coexistence equilibria of the extended model does not exist, however, any potential coexistence equilibrium must satisfy the following system of equations:

\begin{equation}
 \left\lbrace   \begin{split}
        n_* & = \frac{1}{2s}\left(-n_b + \sqrt{n_b^2+4sk(\tilde{\epsilon}\gamma p_* + s\theta)}\right) \\
        z_* &= \frac{1+p_*^2}{p_*}\left(\frac{n_*}{\tilde{k}+n_*}\left(1-\frac{p_*}{c}\right) -\tilde{\epsilon}\right) \\
        d_* &= \frac{1}{\psi}\left((1-\gamma)\frac{p_*}{1+p_*^2} + \tilde{\epsilon}p_*\right) \\
        0 & = \frac{p_*^2}{1+p_*^2} - az_* - \frac{\tilde{\alpha}p_*}{\xi + p_*} \\
    \end{split}\right.
    \label{system}
        \end{equation}
where $$n_b = s(\tilde{k}-\theta) - \tilde{\epsilon}\gamma p_* + \gamma p_*\left(1-\frac{p_*}{c}\right).$$
The positivity of $n_*$ and $d_*$ is clear from their forms provided $p_* > 0$.  Therefore we need to provide a condition such that there exists at least one $p_* > 0$ for which the resulting $z_* > 0$, which as we show below is $\mathcal{P}_0^z > 1$.

\begin{proposition}
For the model \eqref{rescale}, if  $\mathcal{P}_{0}^z > 1$, then zooplankton population is robustly uniformly $\rho$-persistent for $\rho = \min_\tau z(\tau)$ and there exists at least one coexistence equilibrium.
\label{prop:10}
\end{proposition}

\begin{proof}

% First, note that for any coexistence equilibrium of system \eqref{rescale2} $\mathcal{E}_* = (p_*,z_*,n_*,d_*)$ $p_*  \geq c(1-\frac{1}{\mathcal{R}_0^p})$ if and only if \begin{equation}\frac{dp}{d\tau}(p_*) < \epsilon p_*[(\mathcal{R}_0^p - 1) - \mathcal{R}_0^p(p_*/c)] - \frac{p_*^2}{1+p_*^2}z_* < 0.\end{equation}.  Hence $p_* < c(1-1/\mathcal{R}_0^p)$, which provides an upper bound on the set of feasible solutions.  Next, observe that the last equation of system (36) is equivalent to
% \begin{equation}
% \begin{split}
%     z_* &= \frac{p_*}{a}\left(\frac{p_*}{1+p_*^2} - \frac{\tilde{\alpha}}{\xi + p_*}\right) \\
% \end{split} 
% \end{equation}
% Note this implies another set of  conditions on $p_*$, namely 
% \begin{equation}
% \begin{cases}
%      \sqrt{\frac{\alpha}{1-\alpha}} < p_* < c\left(1-\frac{1}{\mathcal{R}_0^p}\right) & \texterm{if } \alpha < \frac{p_*^2}{1+p_*^2} \\
%      \xi >(\alpha + \alpha p_*^2 - p_*^2)/p_* &
% \end{cases}
% \end{equation}
See appendix A.

\end{proof}

We make four key observations about the persistence numbers of zooplankton and phytoplankton which are relevant to our simulations.
\begin{equation}
\left\lbrace\begin{split}
\lim_{\hat{p} \rightarrow c} \mathcal{P}_0^z &= \lim_{\mathcal{P}_0^p \rightarrow\infty} \mathcal{P}_0^z = \frac{1}{\tilde{\alpha}} \\
\lim_{\theta \rightarrow \infty} \mathcal{P}_0^p &= \frac{1}{\tilde{\epsilon}},~ \lim_{\theta\rightarrow 0} \mathcal{P}_0^p = 0 \\
\end{split}\right.
\end{equation}
and further that, for fixed $\mathcal{P}_0^p$,
\begin{equation}
\left\lbrace\begin{split}
\lim_{\tilde{\alpha}\rightarrow\infty} \mathcal{P}_0^z &= \lim_{\xi\rightarrow 0} \mathcal{P}_0^z = 0 \\
\lim_{\tilde{\alpha}\rightarrow 0} \mathcal{P}_0^z &= \lim_{\xi\rightarrow \infty} \mathcal{P}_0^z = \infty 
\end{split}\right.
\end{equation}

\begin{figure}[h!]
\centering 
\begin{subfigure}{\textwidth}
\centering
\includegraphics[width=.5\textwidth]{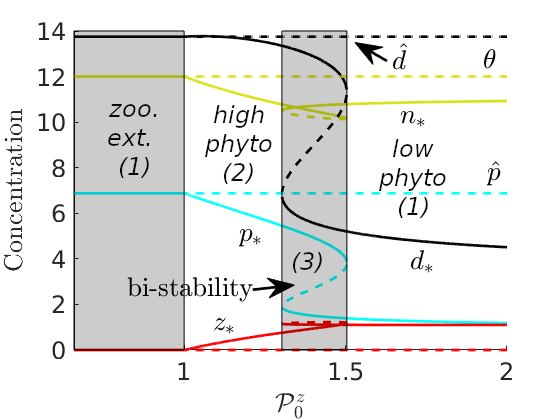}
\caption{}
\end{subfigure}
\begin{subfigure}{.325\textwidth}
\includegraphics[width=\textwidth]{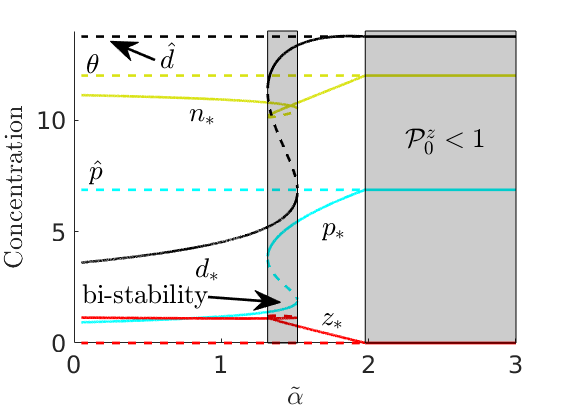}
\caption{}
\end{subfigure}
\begin{subfigure}{.325\textwidth}
\includegraphics[width=\textwidth]{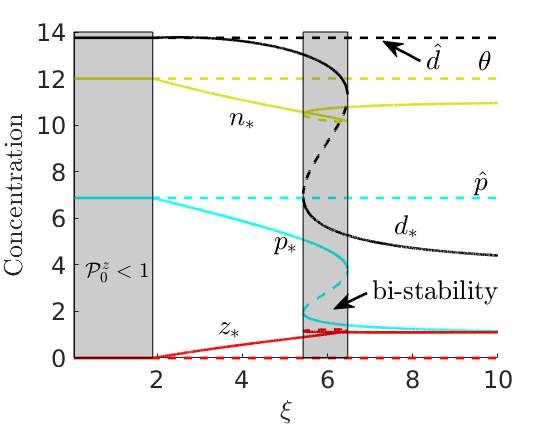}
\caption{}
\end{subfigure}
\begin{subfigure}{.325\textwidth}
\includegraphics[width=\textwidth]{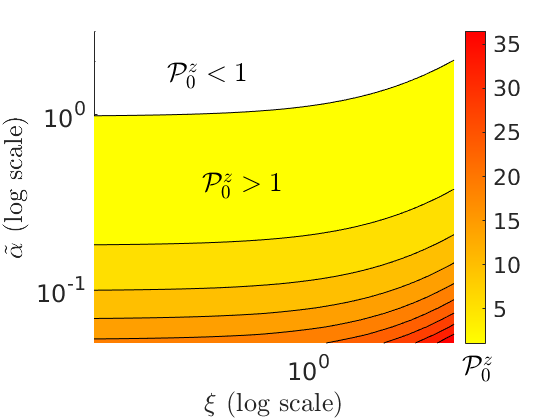}
\caption{}
\end{subfigure}
\begin{subfigure}{.24\textwidth}
\includegraphics[width=\textwidth]{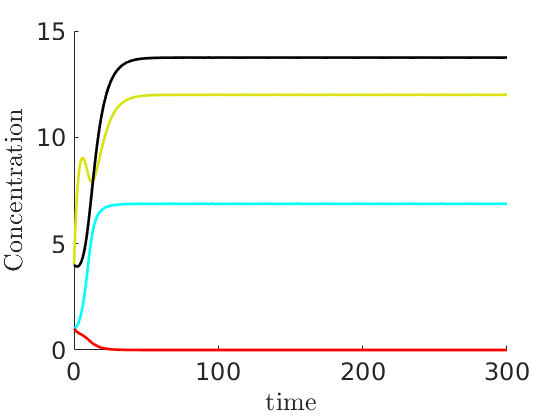}
\caption{}
\end{subfigure}
\begin{subfigure}{.24\textwidth}
\includegraphics[width=\textwidth]{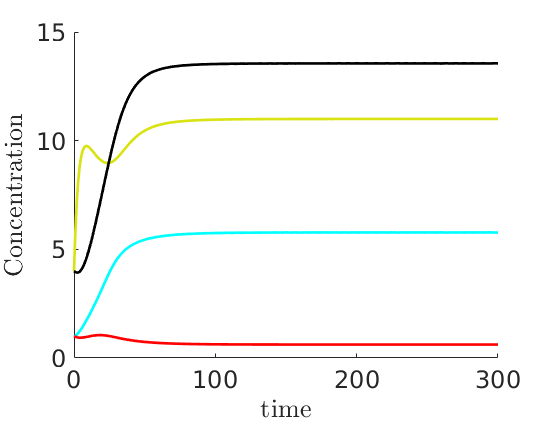}
\caption{}
\end{subfigure}
\begin{subfigure}{.24\textwidth}
\includegraphics[width=\textwidth]{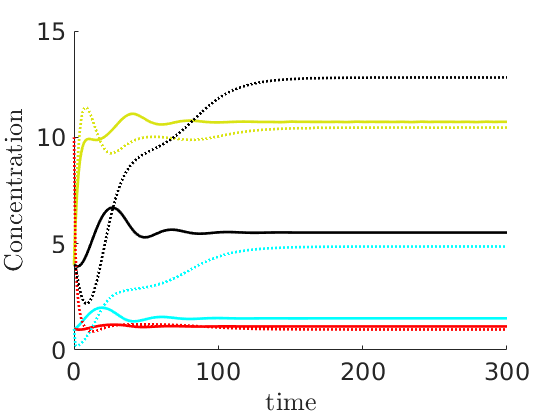}
\caption{}
\end{subfigure}
\begin{subfigure}{.24\textwidth}
\includegraphics[width=\textwidth]{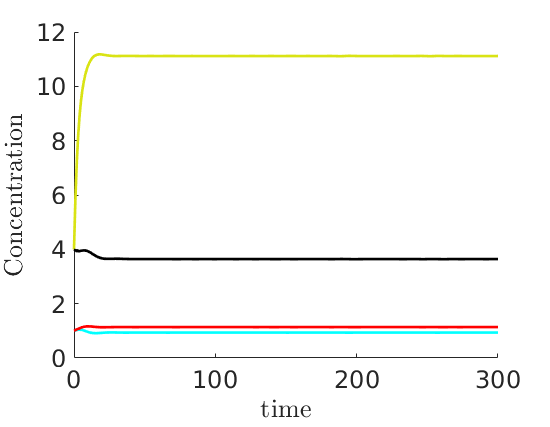}
\caption{}
\end{subfigure}
     \caption{\textbf{Bifurcation diagrams for the rescaled model \eqref{rescale2} with $\tilde{\gamma} = 0.29$ and $\mathcal{P}_0^p \approx 3.2$ during nutrient influx ($\theta = 12$)} when $a = 0.4$.  All other parameter values are in table \ref{tab:sims}.  Blue, red, yellow, and black curves represent phytoplankton, zooplankton, nutrients, and detritus, respectively.  In \textbf{(a)} we present the bifurcation diagram for different values of $\mathcal{P}_0^z$, which is a function of parameters $\tilde{\alpha},\xi$. In \textbf{(b)} and \textbf{(c)} we present one-dimensional bifurcation diagrams in $\tilde{\alpha}$ with $\xi = 7$ and $\xi$ with $\tilde{\alpha} = 1.25$ respectively. Observe that the Phytoplankton population decreases as a function of $\mathcal{P}_0^z.$ In \textbf{(d)}, sensitivity of $\mathcal{P}_0^z$ to variation in $\xi$, $\tilde{\alpha}$ is displayed.  \textbf{(e)-(h)} Time-dependent solutions demonstrating the qualitative dynamics in different parameter regions. The  transition from high to low phytoplankton abundance is region 1 ($\mathcal{P}_0^z = 0.62 \Rightarrow \mathcal{B} = 0.81)$ $\rightarrow$ region 2 ($\mathcal{P}_0^z = 1.24 \Rightarrow \mathcal{B} = 0.61)$ $\rightarrow$ region 3 ($\mathcal{P}_0^z = 1.41 \Rightarrow \mathcal{B} = 0.56)$ $\rightarrow$ region 2  ($\mathcal{P}_0^z = 19.8 \Rightarrow \mathcal{B} = -5.2)$.   }
     \label{fig:bifurcation1}
\end{figure}

\begin{figure}[h!]
\centering
\begin{subfigure}[t]{\textwidth}
\includegraphics[width=\textwidth]{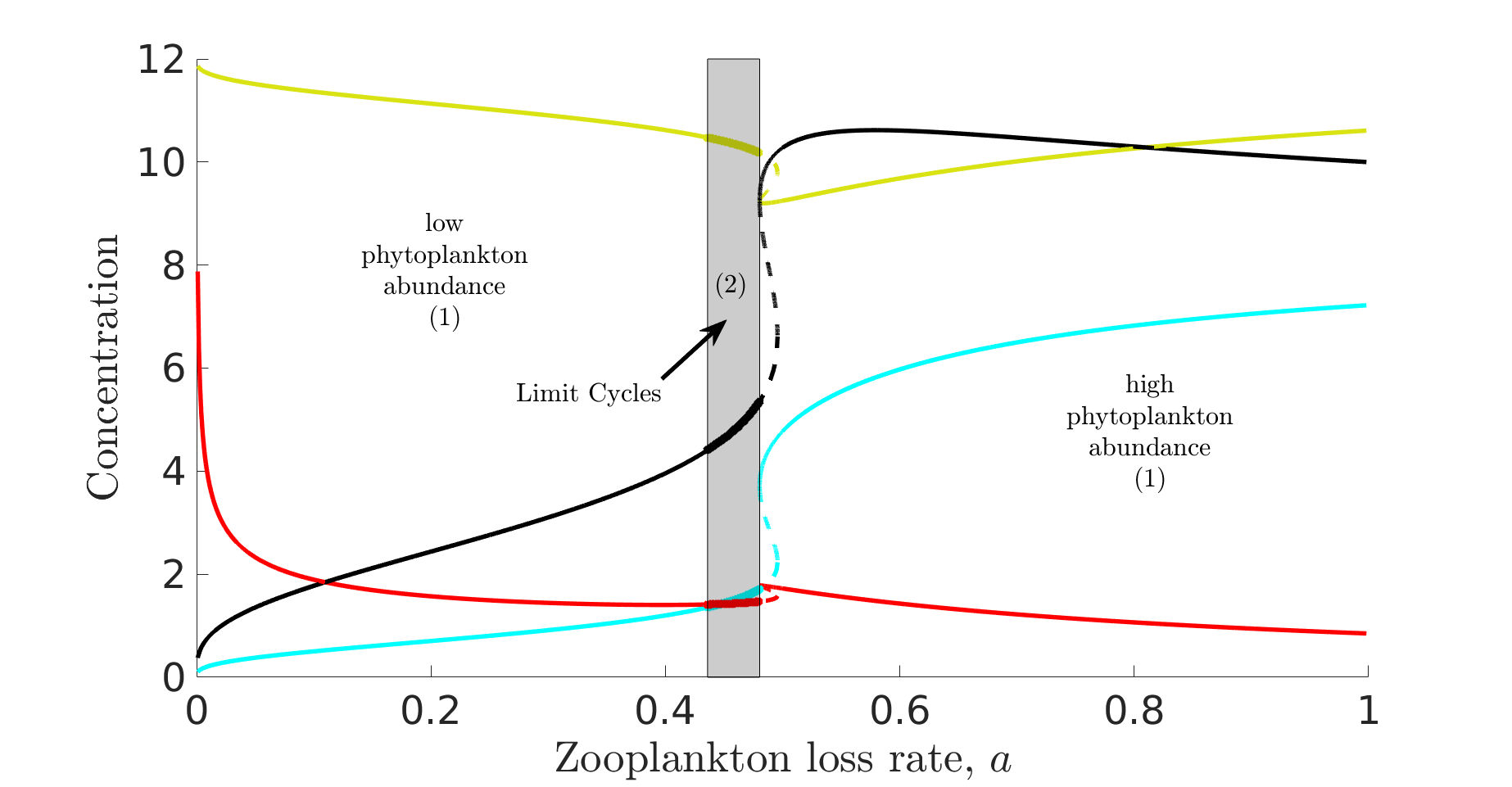}
\caption{}
\end{subfigure}
\begin{subfigure}[t]{.325\textwidth}
\includegraphics[width=\textwidth]{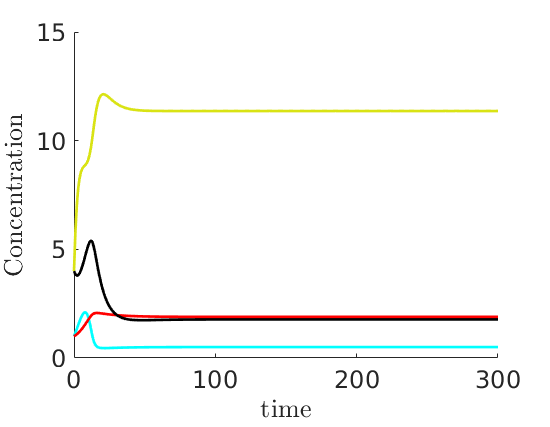}
\caption{}
\end{subfigure}
\begin{subfigure}[t]{.325\textwidth}
\includegraphics[width=\textwidth]{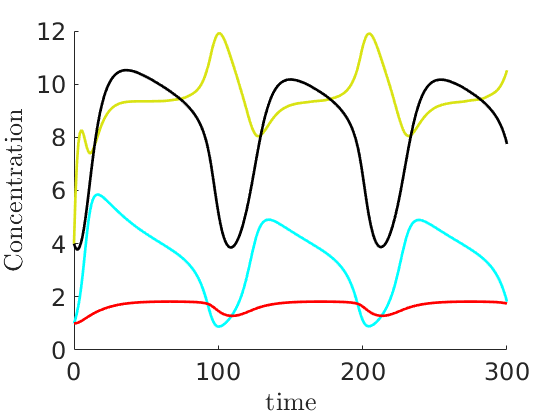}
\caption{}
\end{subfigure}
\begin{subfigure}[t]{.325\textwidth}
\includegraphics[width=\textwidth]{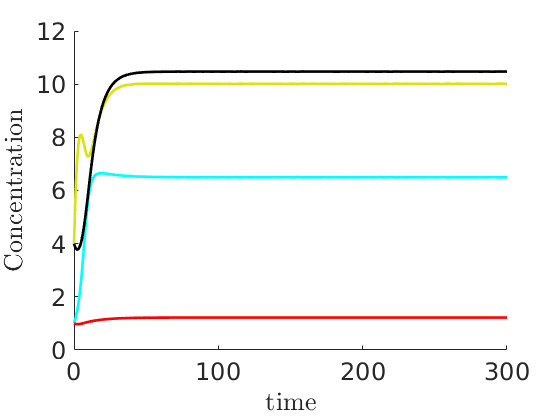}
\caption{}
\end{subfigure}
\caption{One-dimensional bifurcation diagram of the model \eqref{rescale} in $a$ under a parameter regime where the model displays Hopf bifurcation. Notice that, analogous to model \eqref{rescale2} (see figure \ref{fig:quadratic}) the region of stable limit cycles is a transitory state between low and high phytoplankton abundance. Parameter values used are: $\tilde{k} = 0.5 , c = 10, \tilde{\epsilon} = 0.15, \tilde{\gamma} = 0.1 , \tilde{\alpha} = 0.5 , \theta = 12 , \psi = 0.15 , \gamma = 0.5 , \xi = 20$  $(\Rightarrow \mathcal{P}_0^p = 6.4, \mathcal{P}_0^z = 7.65, \mathcal{B} = -0.196)$.  (b)-(d) Time-dependent solutions demonstrating the qualitative dynamics in different parameter regions.  The transition from low to high phytoplankton abundance is region 1 ($a = 0.1$) $\rightarrow$ region 2 ($a = 0.47$) $\rightarrow$ region 1 ($ a = 0.7$).}
\label{fig:FullHopf}
\end{figure}

The NPZD model also presents interesting complex bifurcation dynamics including forward hysteresis (see figure \ref{fig:bifurcation1}).  In the context of population dynamics,  forward hysteresis refers to the appearance of multiple local attractors when a threshold condition (usually a condition analogous to the basic reproduction number, $\mathcal{R}_0$, in disease dynamical systems models) is larger than one \cite{gulbudak2013forward}. For our model specifically, the curve of coexistence equilibria bifurcates from the zooplankton extinction equilibrium when $\mathcal{P}_0^z > 1$.
In (a)-(c), we observe that the curve of phytoplankton equilibria in particular follows a pattern of hysteresis where the transitory state between high and low phytoplankton abundance is a region of bi-stability with model trajectory dependent on initial conditions.  In this region of bi-stability, we note that the basin of attraction for the lower abundance equilibrium is in a neighborhood where the initial phytoplankton and zooplankton populations are approximately equal (see GitHub \cite{MacdonaldCode}) {(d)} sensitivity of $\mathcal{P}_0^z$ to variation in $\xi$, $\tilde{\alpha}$. In Fig. \ref{fig:bifurcation1}(a), we observe that zooplankton will go (locally) extinct when the maximum harmful effect of phytoplankton is more than twice that of the maximum zooplankton grazing rate.  In {(b)}, we see that zooplankton will go (locally) extinct when the square root of the half-saturation constant for grazing is more than twice the half saturation of harmful effect {(c)} phytoplankton population decreases as a function of $\mathcal{P}_0^z$, reflecting the top-down regulation of phytoplankton population by zooplankton predation. For the model \eqref{rescale} Hopf bifurcation can also occur (see figure \ref{fig:FullHopf}). Notice that, similarly to model \eqref{rescale2} (see figure \ref{fig:quadratic}(b)), the region of stable limit cycles is a transitory state between low and high phytoplankton abundance when $\tilde{\gamma}$ is sufficiently small.

\begin{figure}[h!]
\centering
\includegraphics[width=\textwidth]{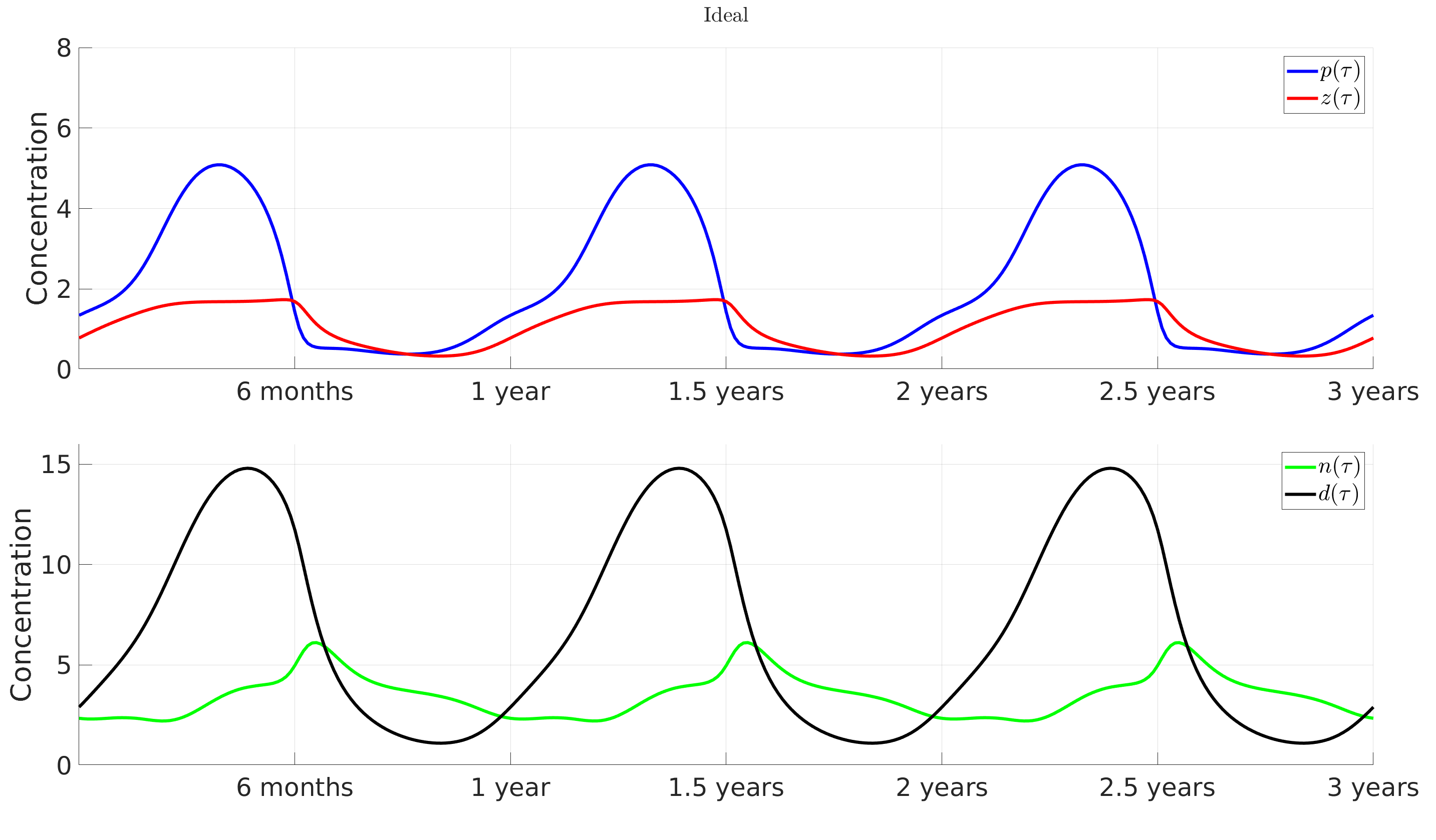}
\caption{\textbf{An ideal healthy ecosystem prior to disturbance} ($\mathcal{B} = 0$) with seasonality incorporated into the time simulations (model \eqref{stochastic}).  The top panel displays plankton populations and the bottom nutrient and detritus levels. $\tilde{k} = 0.5 , c = 10,\tilde{\epsilon} = 0.3, \tilde{\gamma} = 0.29, \tilde{\alpha} = 0.85, \theta = 4, \psi = 0.15,\gamma = 0.5, \xi = 10, a = 0.4, s = 0.3.$ (for the unforced model \eqref{rescale}, $\mathcal{P}_0^p \approx \mathcal{P}_0^z \approx 3)$.}
\label{fig:heal}
\end{figure}

We observe that our model is capable of capturing a wide range of plankton population dynamics as a function of $\mathcal{P}_0^z$ and $\mathcal{P}_0^p$.  To quantify this we define what we term the balance of the ecosystem,
\begin{equation}
    \mathcal{B} := 1 - \frac{\mathcal{P}_0^z}{\mathcal{P}_0^p},
\end{equation}
with the ideal balance being $\mathcal{B} = 0$ (see figures \ref{fig:bifurcation1}, \ref{fig:heal}), and so $\mathcal{P}_0^z = \mathcal{P}_0^p$, and having a positive (negative) value when phytoplankton (zooplankton) are favored.  We observe in particular that our simulations suggest harmful algal blooms, which mathematically we define as a period of time where $dp/d\tau > 0$ and $dz/d\tau < 0$, occur from approximately when $\mathcal{B} > 0.5$ for our choice of other model parameters.  This is chosen because it indicates that despite increasing abundance in the phytoplankton (prey) population,  the zooplankton (predator) population still decreases.  Moreover, zooplankton extinction occurs when $\mathcal{B}$ approaches 1 (see figure \ref{fig:bifurcation1}(e)-(h) and supplementary figures \ref{fig:examples1}-\ref{fig:examples3}).

Thus - as the intrinsic nutrient level of the ecosystem increases in the scenario of a eutrophication event - $\mathcal{P}_0^p$ will increase.  And \emph{if} we are in the region of the parameter space representing an unhealthy ecosystem, so will the size of the region where robust persistence of zooplankton is not assured.  Similarly in the scenario of a re-oligotrophication event as the intrinsic nutrient level of the ecosystem decreases the balance of the ecosystem will change to (temporarily) favor zooplankton -  reflecting the key biological reality we seek to capture with our model.

% \begin{equation}
%   >  p_*(\xi  -  p_*) > \alpha
% \end{equation}
% Next note that for community collapse equilibrium we have Jacobian 
% \begin{equation}
%     J(\mathcal{E}_n') = \begin{pmatrix}
%     \epsilon(\mathcal{R}_0^p-1) & 0 & 0 & 0 \\
%     0 &0 & 0 & 0 \\
%     -\epsilon\mathcal{R}_0^p& 0 & -s & \psi \\
%     \epsilon & 0 & 0 & -\psi
%     \end{pmatrix},
% \end{equation}
% and for this equilibrium we have eigenvalue in the $p$ direction $\lambda_p = \epsilon(\mathcal{R}_0^p-1) > 0$ since $\mathcal{R}_0^p > 1$.  As with Proposition 6, and following the work of \cite{salceanu}, it follows that the system is robustly uniformly persistent when $\min\{\mathcal{R}_0^z, \mathcal{R}_0^p\} > 1$ and that there exists at least one coexistence equilibrium.

% \begin{proof} (Proposition 2)
% For the boundary equilibrium $\E_{np}$ we have the Jacobian:
%   \begin{equation}
%     J(\E_{np}) = \begin{pmatrix} A & 0 \\ * & - s \end{pmatrix},
% \end{equation}
% where $$A = 
% \begin{pmatrix}
% -\frac{\theta}{k+\theta} & -a\mathcal{R}_0 \\
% & \\
% 0 & \tilde{\gamma}a(\mathcal{R}_0 - 1)
% \end{pmatrix},$$
% is block triangular, and $\lambda_z^{\E_{np}} = \tilde{\gamma}a(\mathcal{R}_0 - 1)$.  From this, it is clear that this eigenvalue is strictly positive if $\mathcal{R}_0 > 1$.
% \end{proof}

\subsection{Incorporating seasonality and ecological disturbances}
\begin{figure}[h!]
    \centering
    \begin{subfigure}[t]{.32\textwidth}
    \includegraphics[width=\textwidth]{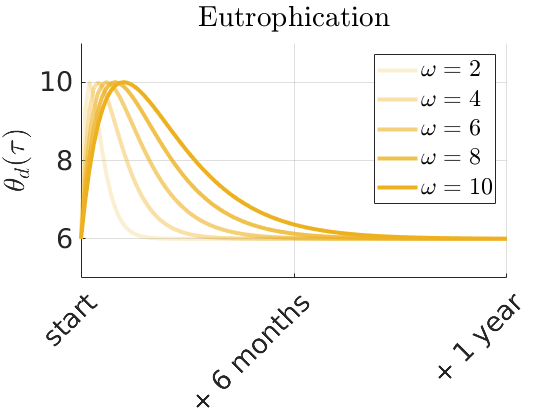}
    \end{subfigure}
        \begin{subfigure}[t]{.32\textwidth}
    \includegraphics[width=\textwidth]{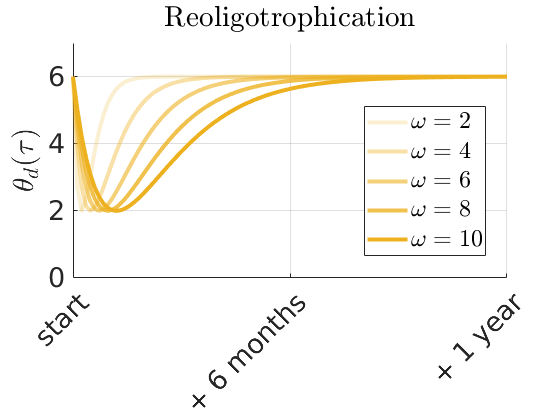}
    \end{subfigure}
     \begin{subfigure}[t]{.32\textwidth}
    \includegraphics[width=\textwidth]{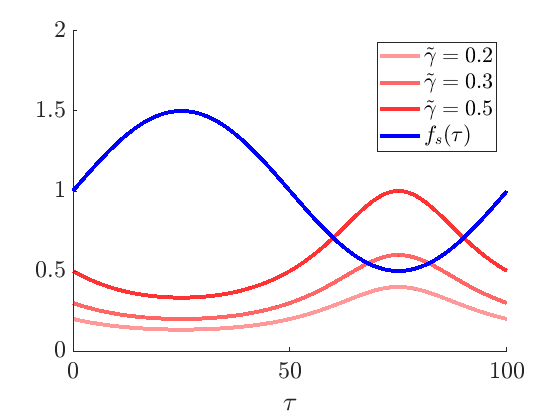}
    \end{subfigure}
    
    \caption{\textbf{Incorporating seasonal variation in light availability and ecological disturbances into the model}.  The duration of the disturbance increases with parameter $\omega$ for disturbance function $\theta_d(\tau,\omega)$ $\textbf{(a-b)}$.  Parameter $M_\theta$ dictates the peak nutrient level during a eutrophication event  \textbf{(a)}, and through nutrient level during a re-oligotrophication event \textbf{(b)}.  In \textbf{(c)} we display seasonality functions $f_s(\tau)$, and $\tilde{\gamma}_s(\tau)$ (with different values of model parameter $\tilde{\gamma}$).}
    \label{fig:dist_func}
\end{figure}

Eutrophication (re-oligotrophication) events may occur at different times, have different peak (trough) nutrient levels, and have different durations; thus here we incorporate seasonality. For a given event, consider the disturbance function, $\theta_d(\tau)$ where  
\begin{equation}
\left\lbrace\begin{split}
    \theta_d(\tau) &= \theta \pm \frac{M_\theta}{\max_\tau g(\omega,\tau)}g(\omega,\tau)  \\
    g(\tau) &= \frac{1}{\omega^2}\tau e^{-\tau/\omega} ,
    \end{split}\right.
\end{equation}

with $M_\theta$ as the maximum intrinsic nutrient level increase/decrease relative to baseline level $\theta$ and $\omega$, a scale parameter, which controls the duration of the disturbance.  We note that this is a modified gamma distribution which is chosen for its flexibility in representing different disturbance curves.  As we illustrate in Fig. \ref{fig:dist_func}, the duration of the disturbance increases with $\omega$. 
Then we define 
\begin{equation}
\theta (\tau) = \begin{cases}
\theta & \tau < \tau_*  \\
\theta_d(\tau) & \tau \geq \tau_*
\end{cases}
\end{equation}
where $\tau_*$ is the disturbance start time.

To incorporate seasonality of light availability we introduce the forcing term $f_s(\tau)$ into our re-scaled model \eqref{rescale}:
\begin{equation}
\left\lbrace\begin{split}
    f_s(\tau) &= 1 + \frac{1}{2}\sin\left(\frac{2\pi\tau}{100}\right) \\
    \tilde{\gamma}_s(\tau) &= \frac{\tilde{\gamma}}{ f_s(\tau)}.
\end{split}\right.~,
\end{equation}
which is the same seasonality function considered in \cite{edwards} and \cite{henderson} (chosen so that the mean value of $f_s$ is one over a complete period).

Finally, putting everything together we arrive at the model,  
\begin{equation}
\left\lbrace
\begin{split}
\frac{dp}{d\tau} &= f_s(\tau)\frac{n}{\tilde{k} + n}p\left(1 - \frac{p}{c}\right) - \frac{p^2}{1 + p^2}z - \tilde{\epsilon}p\\
\frac{dz}{d\tau} &= \tilde{\gamma}_s(\tau)\left[\frac{p^2}{1 + p^2} -\frac{\tilde{\alpha}p}{\xi+p} - az\right]z \\
\frac{dn}{d\tau} &= -\frac{n}{\tilde{k} + n}p\left(1 - \frac{p}{c}\right)  + s(\theta(\tau) - n) + \psi d \\
\frac{dd}{d\tau} &= (1 - \gamma)\frac{p^2}{1 + p^2}z + \tilde{\epsilon}p - \psi d, 
\end{split}
\right.
\label{stochastic}
\end{equation}

A table with a detailed summary of all model terms for the system \eqref{stochastic} is available in the appendices (see table \ref{tab:forms}).
\begin{table}[h!]
\centering
\caption{ Descriptions of model parameters, values, and sources}
\begin{tabular}{lllc}
\hline\noalign{\smallskip}
Param. & Description & Value(s) & Source\\
\noalign{\smallskip}\hline\noalign{\smallskip}
$\tilde{k} $ & Michaelis-Menton half saturation  & 0.5 & \cite{henderson,edwards} \\
$\gamma$ & zooplankton assimilation & 0.5 & \cite{henderson,edwards} \\
$\tilde{\gamma}$ & zooplankton assimilation scaled & 0.29 & \cite{edwards}\\ & by ratio of light response  and & 0.5 & \cite{henderson,edwards}  \\ & grazing rates & (0,1) \\
$c$ & Phytoplankton carrying capacity & 10 & \cite{henderson,edwards}\\
& & (0,15] & \\
$a $ & Zooplankton loss rate & (0,1) & \cite{henderson, edwards}\\
$s$ & nutrient loss/exchange rate & 0.3 & \cite{henderson,edwards}\\
$\theta$ & Intrinsic nutrient level & 4 & \cite{henderson,edwards}\\
& & 12 & \\
$p_0$ & Initial phytoplankton & 1 & \cite{henderson,edwards} \\
& &  [0.375,10] & \\
$z_0$ & Initial zooplankton & 1 & \cite{henderson,edwards} \\
& & [0.1,10] & \\
$n_0$ & Initial nutrients & 4 & \cite{henderson,edwards} \\
& & 1 & \\
$\psi$ & Detritus decay rate & 0.15 & based on \cite{heinle} \\
$\tilde{\epsilon}$ & Phytoplankton natural mortality rate & 0.3 & assumed \\
$\tilde{\alpha}$ & Maximum harmful affect of phytoplankton & (0,3] & -- \\
$\xi$ & Half-saturation of harmful affect & (0,10] & -- \\
$M_\theta$ & Peak/trough nutrient level during disturbance & [0, 12] & -- \\
$\omega$ & Duration of disturbance & [2, 100] & -- \\
\hline\noalign{\smallskip}
\end{tabular}
\label{tab:sims}
\end{table}

\subsection{Ecological implications}
\paragraph{$\mathcal{P}_0^z$ and $\mathcal{P}_0^p$ as measures of ecosystem health}

%  k=p(1); c=p(2); eps=p(3); gtil = p(4); alpha=p(5); theta=p(6); psi=p(7); g = p(8); xi = p(9); a = p(10); s = p(11)
\begin{figure}[h!]
\centering
\includegraphics[width=\textwidth]{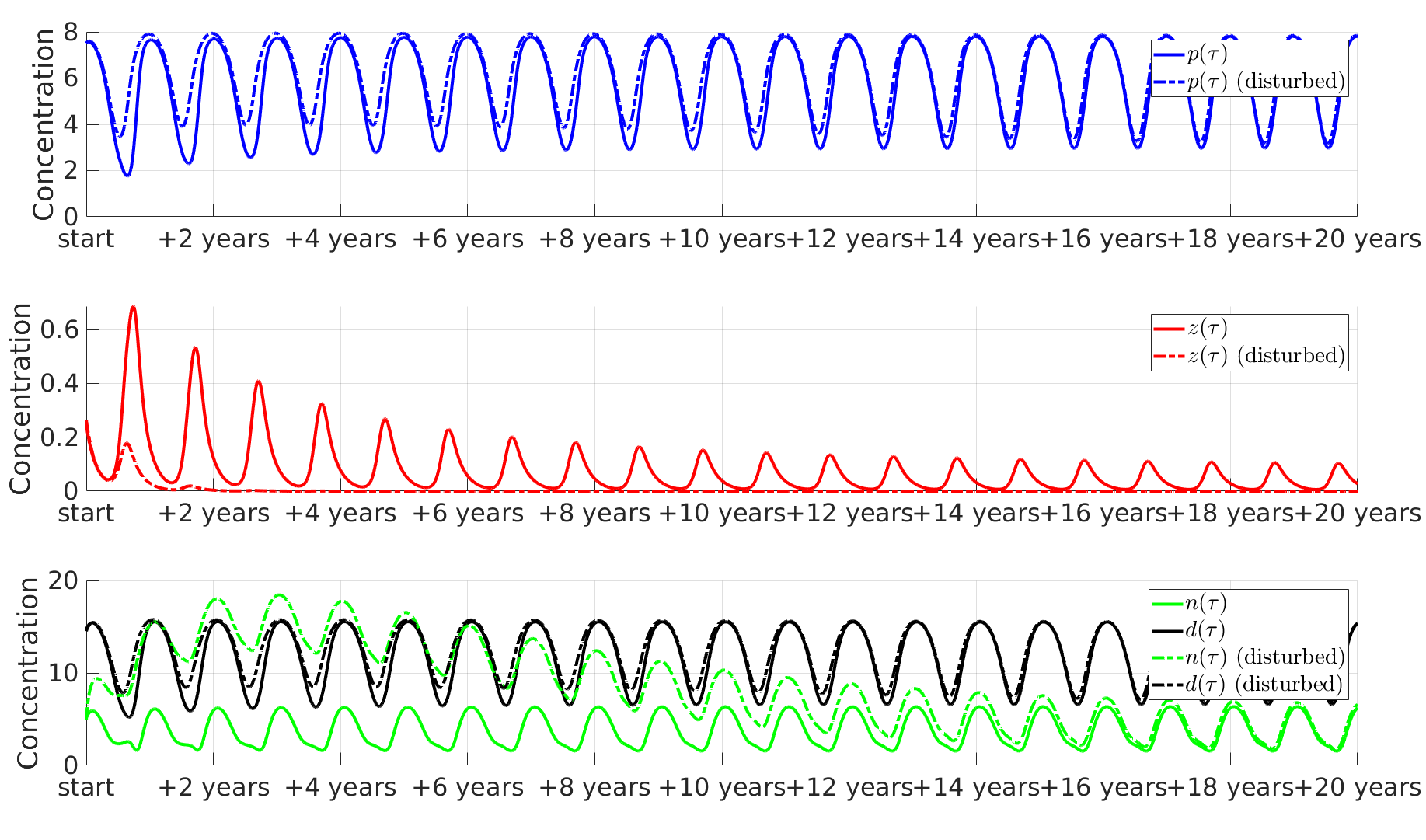}
\caption{\textbf{Eutrophication event in an unhealthy ecosystem a priori favoring phytoplankton}.  Solid lines represent the undisturbed solution and dashed-and-dotted lines the disturbed solution.  If the ecosystem balance, $\mathcal{B}$, sufficiently favors phytoplankton then an influx of nutrients can cause zooplankton extinction (middle panel).  The parameter values used are $\tilde{k} = 0.5 , c = 10,\tilde{\epsilon} = 0.3, \tilde{\gamma} = 0.29, \tilde{\alpha} = 2.81, \theta = 4, \psi = 0.15,\gamma = 0.5, \xi = 10, a = 0.4, s = 0.3, M_\theta = 12, \omega = 308$ (for the unforced model \eqref{rescale}, $\mathcal{P}_0^p \approx 3, \mathcal{P}_0^z \approx 0.9, \mathcal{B} \approx 0.71).$}
\label{fig:eutro}
\end{figure}
\paragraph{Eutrophication}
We further observe that the further out of ideal balance an ecosystem is the more likely that zooplankton will be at risk of (local) extinction during a disturbance (see figure \ref{fig:eutro}) - with an already unhealthy ecosystem which favors phytoplankton being more vulnerable to (additional) eutrophication.  Thus, when ecological factors external to the particular eutrophication event have changed the composition of the existing phytoplankton assemblage there is an elevated risk of this and subsequent knock-on effects from total deregulation of phytoplankton populations with potential far-reaching up-tropic level consequences.

\begin{figure}[h!]
\centering
\includegraphics[width=\textwidth]{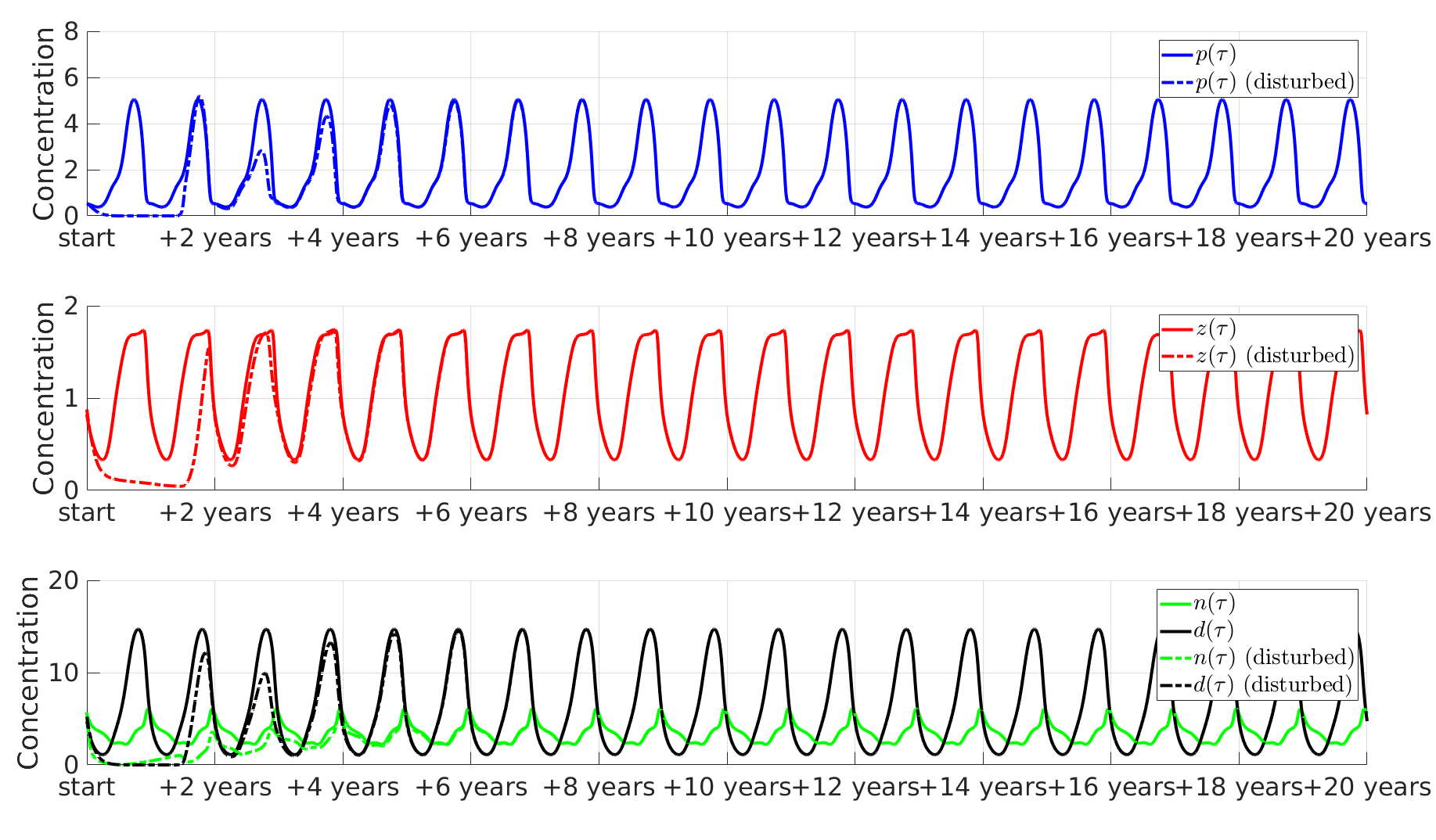}
\caption{\textbf{Re-oligotrophication event in an a priori healthy ecosystem}.  Solid lines represent the undisturbed solution and dashed-and-dotted lines the disturbed solution.  Mathematically, the plankton populations eventually recover to their pre-disturbance dynamics (top and middle panels).  However, in reality, plankton populations may go extinct during the phase at which they are in low concentration, in this simulation approximately from one to three years post-disturbance.  The parameter values used are:  $\tilde{k} = 0.5 , c = 10,\tilde{\epsilon} = 0.3, \tilde{\gamma} = 0.29, \tilde{\alpha} = 0.85, \theta = 4, \psi = 0.15,\gamma = 0.5, \xi = 10, a = 0.4, s = 0.3, M_\theta = 4, \omega = 100$ (for the unforced model \eqref{rescale}, $\mathcal{P}_0^p \approx \mathcal{P}_0^z \approx 3)$.}
\label{fig:reo}
\end{figure}

\paragraph{Reoligotrophication}
In contrast to eutrophication, even a previously healthy ecosystem may be at risk of plankton community collapse provided a prolonged period of re-oligotrophication (see figure \ref{fig:reo}).  Model simulations suggest that even a single year of nutrient depletion is sufficient to cause this collapse.  This may explain the relatively rapid onset of such events in world rivers described by \cite{opposite}.

\section{Discussion}
\label{sec:5}
Starting from a base NPZ modeling framework, we incorporated the harmful effects of phytoplankton overpopulation on zooplankton during HABs, representing a crucial next step in HAB modeling \cite{ralston2020modeling}, and split the nutrient compartment to formulate an NPZD model.  We then mathematically analyze the NPZ system upon which this new model is based - deriving global stability conditions for the zooplankton extinction equilibrium in terms of zooplankton invasion number for a special case of the linear mortality model as well as local stability, Hopf bifurcation, and derived existence conditions for the coexistence equilibria of the NPZ model with both quadratic and linear zooplankton mortality.  We provided one and two-parameter bifurcation diagrams for both models, showing forward hysteresis with bi-stable dynamics and Hopf bifurcation in the quadratic loss case depending on parameter values, and either Hopf bifurcation or transcritical bifurcation in the linear loss case.  Finally, we extended the threshold analysis to the NPZD model, which displays both forward hysteresis with bi-stability and Hopf bifurcation, and examined ecological implications after incorporating seasonality and ecological disturbances in the form of eutrophication and re-oligotrophication events.  Ultimately we quantified ecosystem health in terms of the relative values of the robust persistence thresholds for phytoplankton and zooplankton and found (i) ecosystems sufficiently favoring phytoplankton, as measured by the relative values of plankton persistence numbers, are vulnerable to both HABs and (local) zooplankton extinction (ii) even balanced ecosystems are extremely sensitive to nutrient depletion over relatively short time-scales.

Modeling can provide crucial insights into functional understandings of population dynamics and interactions between ecosystem species and functional groups.  Phytoplankton occupy niches in part based upon water temperature and nutrient composition \cite{barton,chust, richardson}.
Thus phytoplankton niche loss due to anthropologically driven rising water temperatures may cause increased competition in phytoplankton populations and shifts in phytoplankton assemblage composition.  We found that the increasing frequency of harmful algal blooms may be explained, at least in part, by these shifting compositions of phytoplankton assemblages towards types of phytoplankton with more severe harmful effects due to overpopulation as the overall nutrient richness and temperature increase (\cite{hallegraeff1993review}), represented by a decrease in $\mathcal{P}_0^z$ and a shift in balance towards phytoplankton and that this may be exacerbated by eutrophication events, with already unhealthy ecosystems risking (local) zooplankton extinction provided a eutrophication event of sufficient severity occurs (figure \ref{fig:eutro}), representing a tipping point or critical transition \cite{scheffer2001catastrophic,scheffer2009early,scheffer2012anticipating}.

In contrast to eutrophication, we found that even a previously healthy ecosystem is extremely vulnerable to prolonged re-oligotrophication. Model simulations suggest that only one year of nutrient depletion relative to typical intrinsic levels, could cause plankton population collapse. Importantly, original conditions were not recovered by simply reversing the course of nutrient flow showing clear evidence of a tipping point.  This mirrors the perspective of \cite{opposite} and may explain the relatively rapid onset of such events in comparison to eutrophication events (see figure \ref{fig:reo}).

Plankton, by definition, cannot swim against large-scale currents, and of course, live in many ecosystems which are affected by strong currents \cite{karleskint2012introduction}.  Thus, in many settings such as a river, the interplay between biological and physical dynamics is an important factor in the occurrence and severity of harmful algal blooms.  However, a crucial step to understanding the population dynamics resulting from these complex interactions is to first understand bloom dynamics in a simpler physical setting, as we have here. The interplay between physical and biological factors is typically described via a one-way coupling to a diffusion-advection equation with a system of ODEs such as model \eqref{nonrescale} \cite{franks}.  In this framework, the NPZD model can be thought of as the biological dynamics at physical location $x$ and time $t$.  Thus, future work should incorporate fluid dynamics into our modeling framework to understand the interplay between physical and biological factors in the context of ecological disturbances and HABs. Finally, given the potential utility of $\mathcal{P}_0^p$ and particularly $\mathcal{P}_0^z$ as ecosystem monitoring tools - our model should be parameterized to specific ecosystems and specific eutrophication and re-oligotrophication events.

\section*{Acknowledgments}  The authors would like to thank Zachary Topor for informative discussions of plankton ecology.  J.C.M is supported by a Zuckerman Foundation STEM leadership postdoctoral scholarship.  During the main portion of this work, J.C.M. and H.G. were partially supported by a U.S. NSF RAPID grant (no. DMS-2028728) and NSF grant (no. DMS-1951759), and H.G. by a grant from the Simons Foundation/SFARI 638193.
\section*{Data availability statement}
All software for this project is available under a CC-BY-NC 4.0 license and is archived in the Zenodo repository \href{https://doi.org/10.5281/zenodo.7650341}{https://doi.org/10.5281/zenodo.7650341} in citeable format.  

\section*{Appendix}
\setcounter{figure}{0}
\setcounter{table}{0}
\renewcommand{\thefigure}{C\arabic{figure}} 
\renewcommand{\thetable}{B\arabic{table}}
\renewcommand{\thesection}{A\arabic{section}}
\label{sec:Append}
\subsection*{A. Proofs not appearing in the main text}
%c = x(1);
%k = x(2);
%ghat = x(3);
%a = x(4);
%g = x(5);
%s = x(6);
%n0 = x(7);
%u0 = x(8:10);
  %x = [10,.5,.1,.37,.29,.3,12,P(j,k),Z(j,k),1]; % limit cycle
        %x = [10,.5,.5,.41,.29,.3,12,P(j,k),Z(j,k),1]; % bi-stability
        %x = [10,.5,.5,.2,.29,.3,12,P(j,k),Z(j,k),1]; % zoo dominates 
  %      x = [10,.5,.29,.6,.29,.3,12,P(j,k),Z(j,k),1]; % phyto dominates 
% \begin{table}[h!]
% \caption{Parameter values used in simulations and bifurcation diagrams}
%     \centering
%     \begin{tabular}{cccc}
%     \hline 
%     parameter & value(s) & Source  \\
%     \hline 
%      $c$    & 10 & \cite{henderson, edwards}\\
%      & (0, 15] & \\
%       $\tilde{k}$ & 0.5 &
%       \cite{henderson,edwards} \\
%       $\tilde{\gamma}$ & 0.5 & \cite{edwards} \\
%       & 0.29 & \cite{henderson,edwards} \\
%       & (0, 1) & \\
%       $a$ & (0, 1) & \cite{henderson,edwards} \\
%       $\gamma$ & 0.5 &  \cite{henderson,edwards}\\
%       $s$ & 0.3 & \cite{henderson,edwards} \\
%       $\theta$ & 4 & \cite{henderson,edwards} \\
%       & 12 & \\
%       $z_0$ & [0.1,10] & \\
%       $p_0$ & [0.375,4] & \\
%       $n_0$ & 1 & \\ 
%       & 4 & \cite{henderson,edwards} \\
%       \end{tabular}
%     \label{tab:my_label}
% \end{table}

\begin{proof} (Prop. \ref{prop:append})
Consider the Jacobian matrix evaluated at this point, which is \begin{equation}
    J(\E_n) = \begin{pmatrix}
		\frac{\theta}{\tilde{k} + \theta} & 0 & 0\\
		\\
		0 & -\tilde{\gamma}a(1-\sigma) & 0 \\ \\
		-\frac{\theta}{\tilde{k} + \theta} & 0 & -s
	\end{pmatrix}.
	\label{last}
\end{equation}
From the form of \eqref{last} it is readily apparent that $\E_n$ is always unstable. \\
\end{proof}
% \begin{proof} (Proposition 2)
% For the boundary equilibrium $\E_{np}$ we have the Jacobian:
%   \begin{equation}
%     J(\E_{np}) = \begin{pmatrix} A & 0 \\ * & - s \end{pmatrix},
% \end{equation}
% where $$A = 
% \begin{pmatrix}
% -\frac{\theta}{k+\theta} & -a\mathcal{R}_0 \\
% & \\
% 0 & \tilde{\gamma}a(\mathcal{R}_0 - 1)
% \end{pmatrix},$$
% is block triangular, and $\lambda_z^{\E_{np}} = \tilde{\gamma}a(\mathcal{R}_0 - 1)$.  From this, it is clear that this eigenvalue is strictly positive if $\mathcal{R}_0 > 1$.
% \end{proof}
\begin{proof} (\hypertarget{foo1}{Prop. \ref{prop:1}})
Note that when $\sigma = 0$, the Jacobian evaluated at equilibrium $\E_{np}$ takes the form of 

  \begin{equation}
    J(\E_{np}) = 
    \begin{pmatrix} 
    A & 0 \\ * & - s 
    \end{pmatrix},
\end{equation}
where $$A = 
\begin{pmatrix}
-\frac{\theta}{k+\theta} & -a\mathcal{R}_0 \\
& \\
0 & \tilde{\gamma}a(\mathcal{R}_0-1)
\end{pmatrix}.$$
\end{proof}

 \begin{proof} (\hypertarget{foo2}{Prop. \ref{prop:3}}) \hypertarget{foo1}{First}, note that given system \eqref{rescale} and $\sigma = 0$ the Jacobian matrix evaluated at $\E_*$ is,
\begin{equation}J(\E_*) =
    \begin{pmatrix}
    \psi_1 + \psi_2 - \psi_3 - \psi_4 & & & -a & & & \psi_5 \\
    &  & & & & & \\
    \tilde{\gamma}z_*(\psi_3 - \psi_2) & & & 0 & & & 0 \\
    & & & & & & \\
    -\psi_1 + (1-\gamma)(\psi_3-\psi_2) +  \psi_4 & & & a(1-\gamma) & & & -\left(\psi_5+s\right)
    \end{pmatrix}
\end{equation}
It follows that we have characteristic polynomial (assisted by MatLab Symbolic Math Toolbox): \begin{equation}\chi_{J(\E_*)}(\lambda) = \lambda^3 + \xi_2\lambda^2 + \xi_1\lambda + \xi_0.\end{equation}  Thus we have Hurwitz determinants, $H_i$:
\begin{equation}
    \begin{split}
        H_1 & = \xi_2 \\
        H_2 & = \xi_2\xi_1 - \xi_0 \\
        H_3 & = \xi_0H_2
    \end{split}
\end{equation}

The generalized Routh-Hurwitz criterion indicates that $\E_*$ will be locally asymptotically stable if and only if $H_i > 0$ $\forall i$.  This is equivalent to condition \eqref{linearCoStab}.  Liu \cite{liu} also indicates that if \begin{equation}H_1 > 0, H_2\vert_{\alpha_0} = 0,\textrm{ and }\frac{d}{d\alpha}H_2\vert_{\alpha_0} \neq 0\end{equation} thus a necessary condition for a simple Hopf bifurcation to occur is $\xi_0 = \xi_1\xi_2$.  Note condition (57) is equivalent to conditions \eqref{linearHopf1}, \eqref{linearHopf2}. 
\end{proof}

\begin{proof} (\hypertarget{foo3}{Prop. \ref{prop:special}})
Existence is clear from the form of $\E_c$.  Note that
\begin{equation}
    J(\E_{c}) = \begin{pmatrix} 
-\frac{(N_T-c)}{\tilde{k} + N_T - c} & * \\
0 & \gamma a (\mathcal{R}_0 - 1) 
\end{pmatrix}.
\end{equation}
\end{proof}

\begin{proof} (\hypertarget{foo4}{Prop. \ref{prop:5}}) 
\hypertarget{foo2}{The} proof is similar to that of Prop. \ref{prop:3} with the difference in detail due only to the Jacobian in the quadratic loss case being:
\begin{equation}J(\E_*) =
    \begin{pmatrix}
    \psi_1 + \psi_2 - \psi_3 - \psi_4 & & & -az_* & & & \psi_5 \\
    &  & & & & & \\
    \tilde{\gamma}z_*(\psi_3 - \psi_2) & & & -a\tilde{\gamma}z_* & & & 0 \\
    & & & & & & \\
    -\psi_1 + (1-\gamma)(\psi_3-\psi_2)  + \psi_4 & & & az_*(1-\gamma) & & & -\left(\psi_5 + s\right)
    \end{pmatrix}
\end{equation}
and so our characteristic polynomial is
\begin{equation}\chi_{J(\E_*)}(\lambda) = \lambda^3 + \nu_2\lambda^2 + \nu_1\lambda + \nu_0.\end{equation}.
\end{proof}

\begin{proof} (\hypertarget{foo5}{Prop. \ref{prop:8}})
The proof of dispatavity of model \eqref{rescale} is similar to the proof for such in model \eqref{rescale2} (proposition \ref{prop:4}).  Observe that the Jacobian evaluated at this equilibrium is 
\begin{equation}
J(\mathcal{E}_n') = \begin{pmatrix}
		\tilde{\epsilon}(\mathcal{P}_0^p-1) & 0 & 0 & 0\\
		\\
		0 & 0 & 0 & 0\\ \\
		-\frac{\theta}{\tilde{k} + \theta} & 0 & -s & \psi \\ \\
\tilde\epsilon & 0 & 0 & -\psi
	\end{pmatrix}
\end{equation}
which is block triangular and has eigenvalue $\tilde{\epsilon}(\mathcal{P}_0^p-1$) in the $p$ direction.  Thus via \cite{salceanu} as in our earlier proofs the desired result follows.
\end{proof}

\begin{proof} (\hypertarget{foo6}{Prop. \ref{prop:10}})
First, we observe that since $\mathcal{P}_0^z > 1$ phytoplankton, upon which zooplankton depend, are robustly persistent, and the zooplankton equilibrium $\mathcal{E}_{npd}$ exists.  Next, notice that since a necessary condition for $\mathcal{P}_0^z > 1$ is $\mathcal{R}_{0,1}^z > 1$ it follows that the zooplankton extinction equilibrium is unstable.  Next, observe that the eigenvalue in the $z$ direction for this equilibrium is 
\begin{equation}
    \lambda_z^{\mathcal{E}_{npd}} = b_1(\mathcal{R}_{0,1}^z-1),
\end{equation}
(see equation \eqref{npdJac}), which is positive whenever $\mathcal{P}_0^z > 1$.  Thus as previously (by the work of \cite{salceanu}) zooplankton are robustly persistent, and it follows that there exists at least one coexistence equilibrium.
\end{proof}
\clearpage
\subsection*{B. Supplementary tables}

\begin{table}[h!]
\centering
\caption{Summary of specific functional forms for the re-scaled system incorporating seasonality and ecological disturbances (model \eqref{stochastic})}
\begin{tabular}{ll}
\hline\noalign{\smallskip}
Functional Form & Description \\
\noalign{\smallskip}\hline\noalign{\smallskip}
$1 + \frac{1}{2}\sin(2\pi\tau/100)$ & $f_s(\tau)$, Seasonal light availability\\
& \\
$\dfrac{n}{\tilde{k} + n} $ & Nutrient uptake rate \\
& \\
$\left(1-\dfrac{p}{c}\right) $ & Phytoplankton reproduction rate\\
& \\
$\tilde{\epsilon}$ & Phytoplankton mortality rate \\
& \\
$\dfrac{p^2}{1+p^2}$ & Zooplankton grazing rate\\
& \\
$\tilde{\gamma}/f_s(\tau)$ & $\tilde{\gamma}_s(\tau)$, zooplankton assimilation  rate \\
&\\
$\dfrac{p^2}{1+p^2}-\dfrac{\tilde{\alpha}p}{\xi+p}$ & $r(p)$, net zooplankton growth rate \\
& \\
$a z$ & Zooplankton mortality rate\\
& \\
$\begin{cases}  \theta_d(\tau)&= \theta \pm \frac{M_\theta}{\max_\tau g(\omega,\tau)}g(\omega,\tau)  \\
    g(\tau) &= \frac{1}{\omega^2}\tau e^{-\tau/\omega} ,
    \end{cases}$ & Disturbance function\\
    & \\$\theta (\tau) = \begin{cases}
\theta & \tau < \tau_*  \\
\theta_d(\tau) & \tau \geq \tau_*
\end{cases}$  & Intrinsic nutrient level\\
& \\
$\psi$ & Nutrient decay rate \\
\bottomrule
\end{tabular}
\label{tab:forms}
\end{table}
\clearpage
\subsection*{C. Supplementary figures}

\begin{figure}[h!]
\centering 
 \begin{subfigure}[t]{.24\textwidth}
     \includegraphics[width=\textwidth]{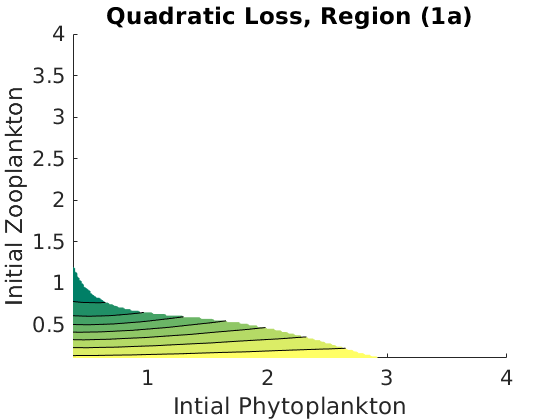}
     \caption{$a = 0.2$}
     \end{subfigure}
     \begin{subfigure}[t]{.24\textwidth}
     \includegraphics[width=\textwidth]{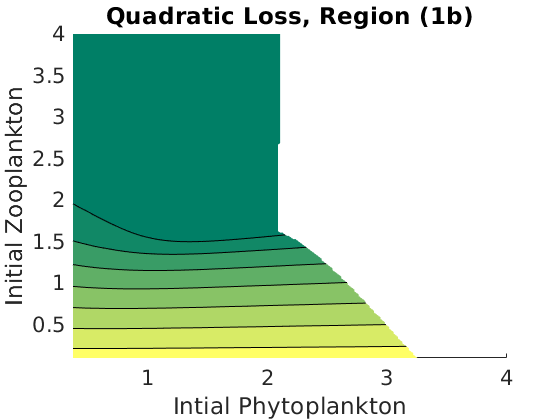}
     \caption{$a = 0.37$}
     \end{subfigure}
     \begin{subfigure}[t]{.24\textwidth}
     \includegraphics[width=\textwidth]{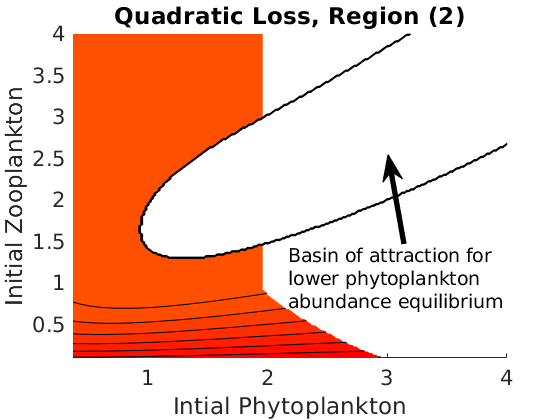}
     \caption{$a = 0.41$}
     \end{subfigure}
     \begin{subfigure}[t]{.24\textwidth}
     \includegraphics[width=\textwidth]{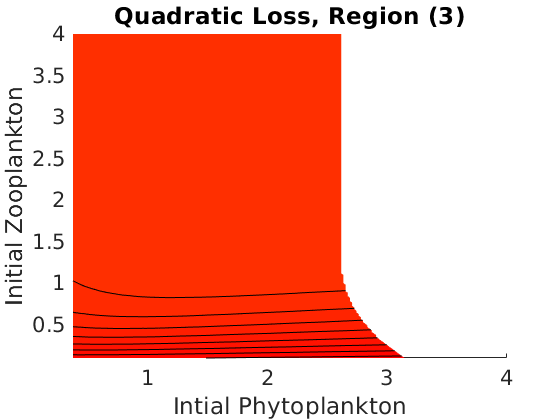}
     \caption{$a = 0.6$ }
     \end{subfigure}
     \caption{Bloom occurrence as a function of plankton population at the moment of maximum eutrophication ($\theta = 12$) in the NPZ subsystem at different zooplankton mortality rates using quadratic zooplankton loss term.  While other model parameters can change qualitative dynamics, zooplankton loss rate, $a$ is the primary driver of likelihood and type of bloom occurrence, with an implied critical threshold beyond which when a bloom occurs it will be harmful.  All other parameter values used are provided in table \ref{tab:sims}.}
     \label{fig:bloomQuad}
\end{figure}

\begin{figure}[h!]
\centering 
 \begin{subfigure}[t]{.24\textwidth}
     \includegraphics[width=\textwidth]{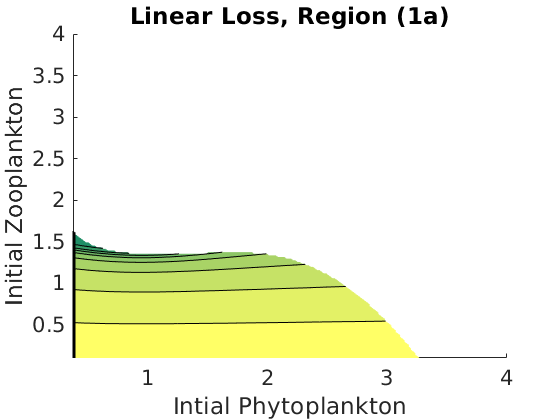}
     \caption{$a = 0.37$}
     \end{subfigure}
     \begin{subfigure}[t]{.24\textwidth}
     \includegraphics[width=\textwidth]{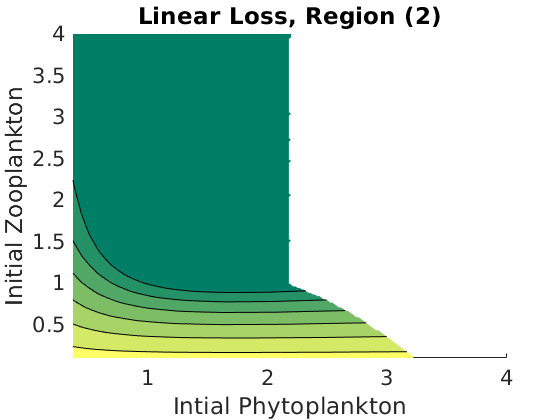}
     \caption{$a = 0.7$}
     \end{subfigure}
     \begin{subfigure}[t]{.24\textwidth}
     \includegraphics[width=\textwidth]{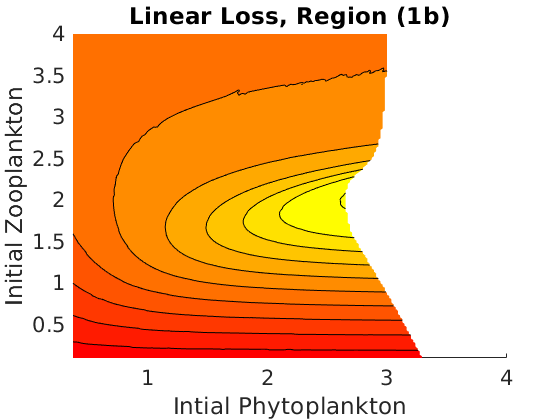}
     \caption{$a = 0.98$}
     \end{subfigure}
     \begin{subfigure}[t]{.24\textwidth}
     \includegraphics[width=\textwidth]{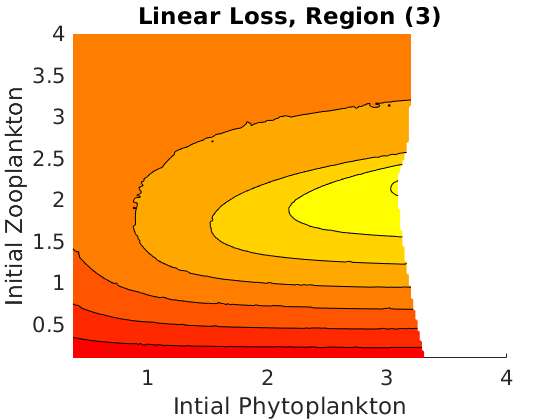}
     \caption{$a = 0.999$ }
     \end{subfigure}
     \caption{Bloom occurrence as a function of plankton population at the moment of maximum eutrophication ($\theta = 12$) in the NPZ subsystem at different zooplankton mortality rates using linear zooplankton loss term.  The primary difference from the quadratic loss model (Fig. \ref{fig:bloomQuad}) is a much larger zooplankton loss rate to induce different bloom dynamics.  While other model parameters can change qualitative dynamics, zooplankton loss rate, $a$ is the primary driver of likelihood and type of bloom occurrence, with an implied critical threshold beyond which when a bloom occurs it will be harmful.  All other parameter values used are provided in table \ref{tab:sims}.}
     \label{fig:bloomLinear}
\end{figure}
\begin{figure}[h!]
    \centering
    \includegraphics[width=\textwidth]{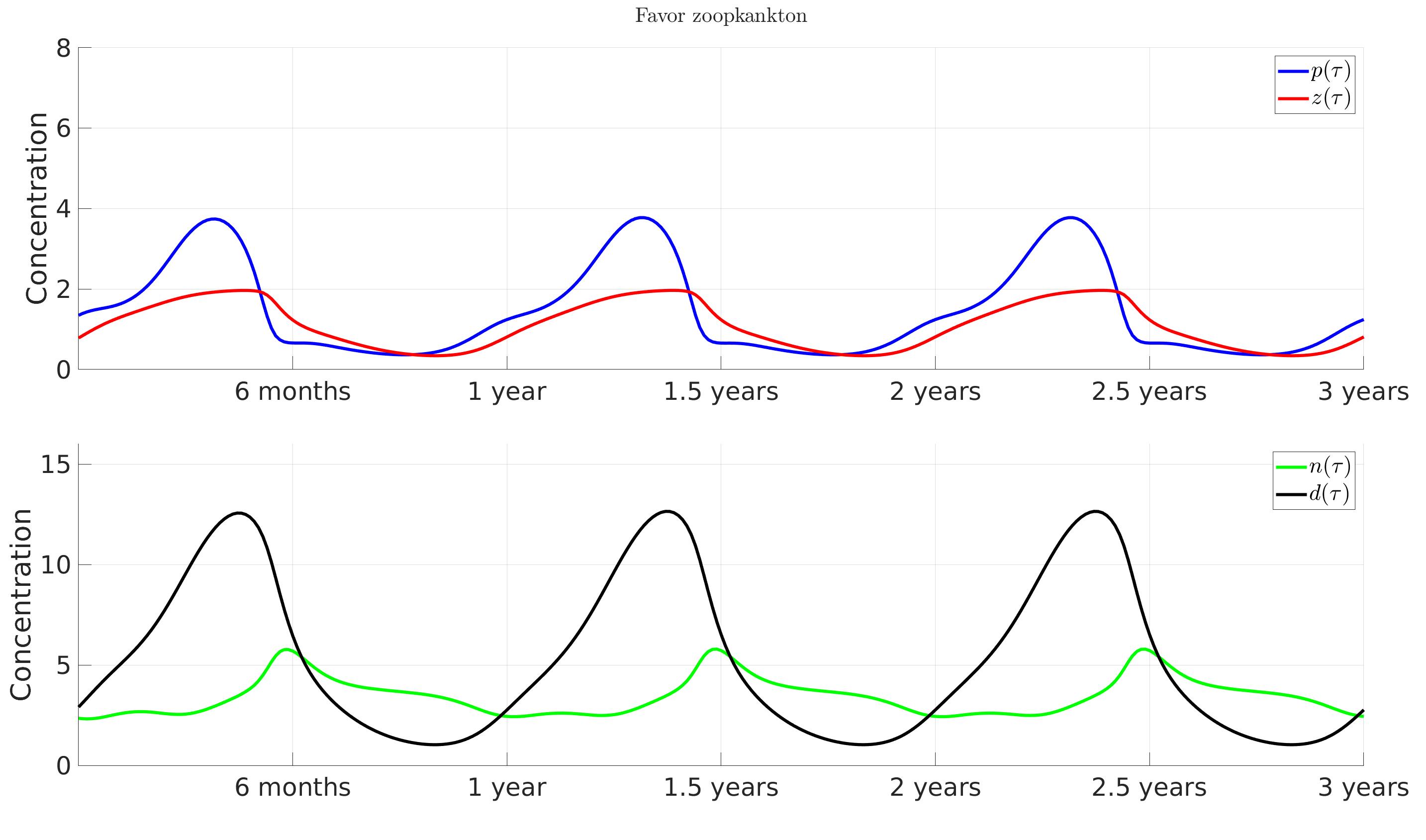}
  \caption{The model dynamics as a function of ecosystem balance after incorporating seasonality in the region of the parameter space where ecosystem balance favors zooplankton.  The parameter values used are: $\tilde{k} = 0.5 , c = 10,\tilde{\epsilon} = 0.3, \tilde{\gamma} = 0.29, \tilde{\alpha} = 0.5, \theta = 4, \psi = 0.15,\gamma = 0.5, \xi = 10, a = 0.4, s = 0.3$ (for the unforced model \eqref{rescale}, $\mathcal{P}_0^p \approx 3, \mathcal{P}_0^z \approx 5, \mathcal{B} = -0.67)$.}
  \label{fig:examples1}
\end{figure}
\begin{figure}[h!]
    \centering
    \includegraphics[width=\textwidth]{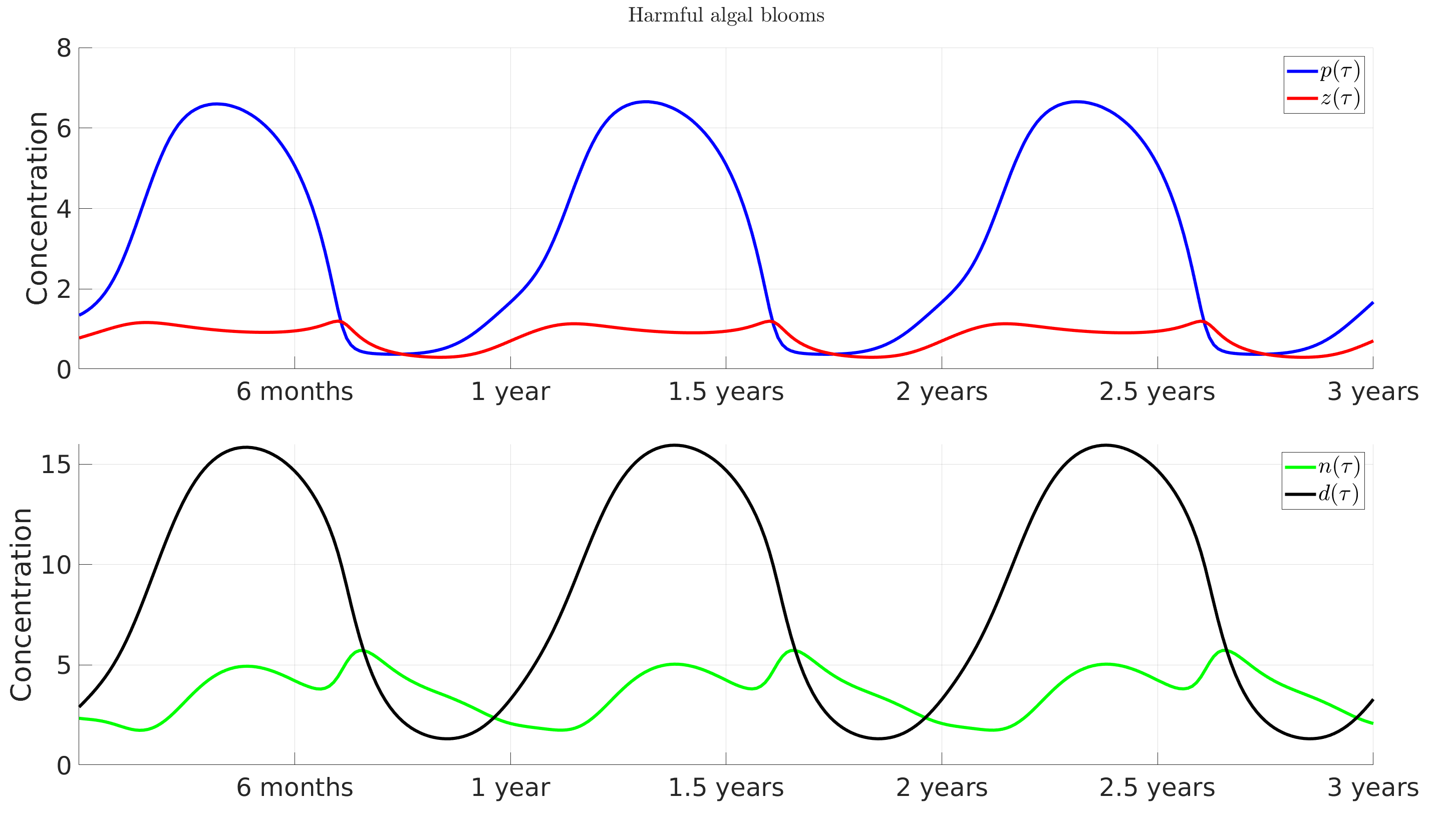}
  \caption{The model dynamics as a function of ecosystem balance after incorporating seasonality in the region of the parameter space where ecosystem balance favors phytoplankton and HABs occur. The parameter values used are: $\tilde{k} = 0.5 , c = 10,\tilde{\epsilon} = 0.3, \tilde{\gamma} = 0.29, \tilde{\alpha} = 1.6, \theta = 4, \psi = 0.15,\gamma = 0.5, \xi = 10, a = 0.4, s = 0.3$ (for the unforced model \eqref{rescale}, $\mathcal{P}_0^p \approx 3, \mathcal{P}_0^z \approx 1.5, \mathcal{B} = 0.48)$.}
  \label{fig:examples2}
\end{figure}
\begin{figure}[h!]
    \centering
    \includegraphics[width=\textwidth]{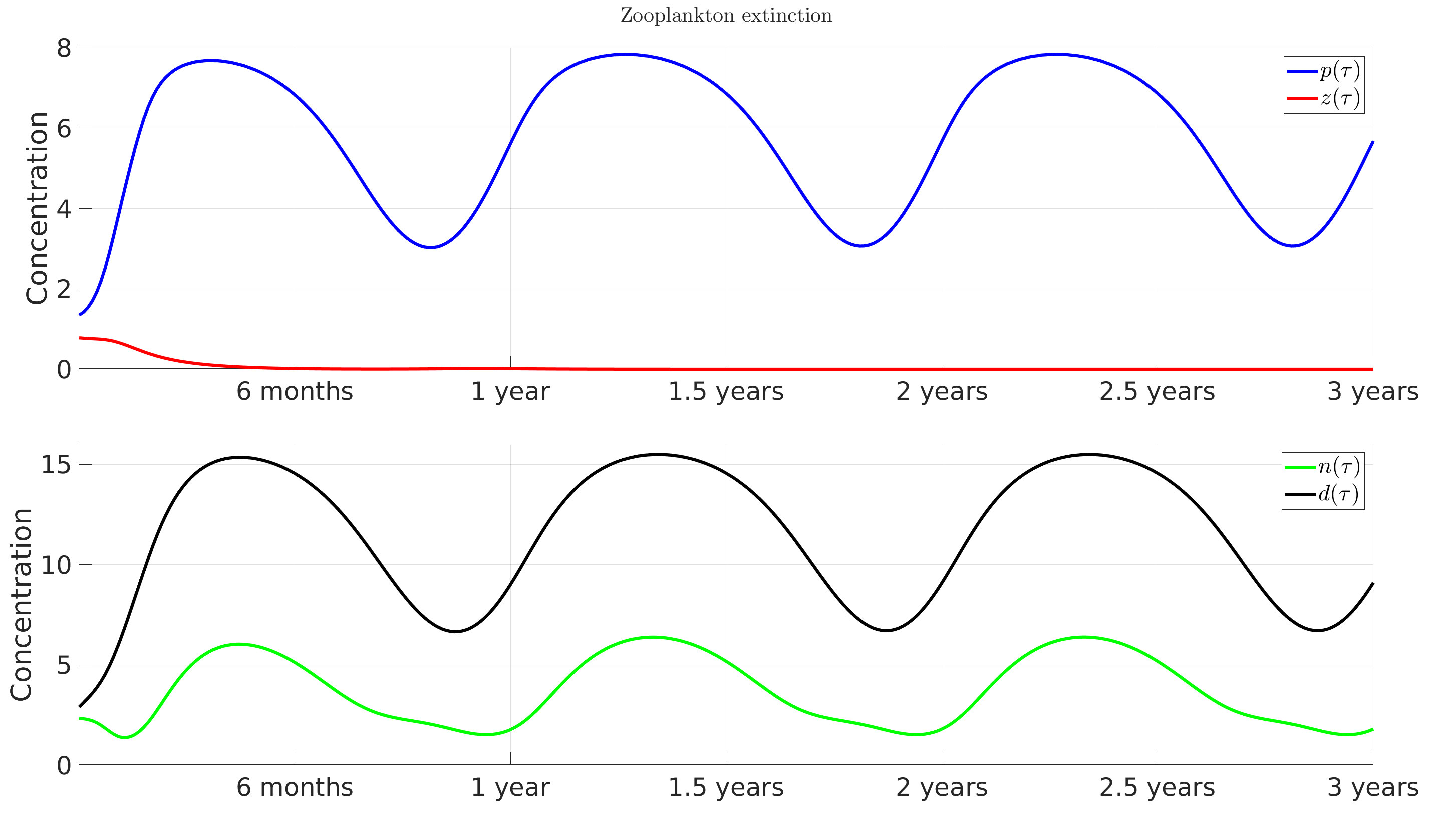}
  \caption{The model dynamics as a function of ecosystem balance after incorporating seasonality in the region of the parameter space where ecosystem balance favors phytoplankton and zooplankton extinction occurs. The parameter values used are: $\tilde{k} = 0.5 , c = 10,\tilde{\epsilon} = 0.3, \tilde{\gamma} = 0.29, \tilde{\alpha} = 3.2, \theta = 4, \psi = 0.15,\gamma = 0.5, \xi = 10, a = 0.4, s = 0.3$ (for the unforced model \eqref{rescale}, $\mathcal{P}_0^p \approx 3, \mathcal{P}_0^z \approx 1.5, \mathcal{B} = 0.74)$.}
  \label{fig:examples3}
\end{figure}

\clearpage
%\end{document}
% end of file template.tex



\bibliography{Revision.bib}
\end{document}